\documentclass[preprint,aps,showpacs,preprintnumbers,amsmath,amssymb]{revtex4}

\usepackage{graphicx}
\usepackage{dcolumn}
\usepackage{bm}

\begin{document}


\title{Quantum number projection at finite temperature
via thermo field dynamics}

\author{K. Tanabe}
\email{tanabe@phy.saitama-u.ac.jp}
\affiliation{Department of Physics, Faculty of Science, 
Saitama University, Sakura, Saitama 338-8570, Japan}
\author{H. Nakada}
\email{nakada@faculty.chiba-u.jp}
\affiliation{Department of Physics, Faculty of Science, 
Chiba University, Inage, Chiba 263-8522, Japan}

\date{\today}

\begin{abstract}
Applying the thermo field dynamics, 
we reformulate exact quantum number projection 
in the finite-temperature Hartree-Fock-Bogoliubov theory. 
Explicit formulae are derived for the simultaneous projection 
of particle number and angular momentum, 
in parallel to the zero-temperature case. 
We also propose a practical method
for the variation-after-projection calculation,
by approximating entropy 
without conflict with the Peierls inequality. 
The quantum number projection 
in the finite-temperature mean-field theory 
will be useful to study effects of quantum fluctuations 
associated with the conservation laws on thermal properties of nuclei. 
\end{abstract}

\pacs{21.60.Jz, 21.10.-k, 21.90.+f, 05.30.Fk}

\keywords{Quantum number projection,
Finite-temperature Hartree-Fock-Bogoliubov theory,
Thermo field dynamics}
\maketitle

\section{INTRODUCTION}

Transitions among different phases have been observed
as temperature increases or decreases,
in wide variety of physical systems
from elementary particles to condensed matters.
They have attracted great interest
both experimentally and theoretically,
and profound physics content has been revealed
particularly in infinite systems.
Despite an isolated system,
statistical properties of atomic nuclei have been investigated
in terms of the so-called nuclear temperature~\cite{BM1},
providing us with a ground
to study phase transitions in nuclei.
However, it is not straightforward to identify phase transitions
in finite systems like nuclei,
because the quantum fluctuations
associated with finiteness of the system
obscure the transitions~\cite{LA84,NA97}
so that even definition of `phases' should
depend on theoretical models to a considerable extent.

It will be natural, from theoretical points of view,
to define phases in nuclei via the mean-field picture,
since `phases' in physical systems
are more or less a semi-classical concept.
Indeed, we discuss spherical/deformed
and normal-fluid/superfluid phases in nuclei
based on the mean-field approximation.
Phase transitions are often connected
to the spontaneous symmetry breaking,
leading to violation of conservation laws.
In the mean-field picture of nuclei,
the rotational symmetry is broken in the deformed phase,
and the particle number conservation is violated
in the superfluid phase.
However, the symmetry breaking in nuclei is a hypothetical effect
due to the mean-field approximation.
In practice, the conservation laws are restored
via a sort of quantum fluctuations
arising from correlations beyond the mean field.
In studying statistical properties of finite systems,
it will be significant to investigate effects of the symmetry 
restoration. 

As the most general mean-field theory, we shall primarily consider 
the Hartree-Fock-Bogoliubov (HFB) approximation. 
To restore the symmetries, projection with respect to the 
quantum numbers should be applied~\cite{RS80}. 
The particle number projection 
as well as the angular momentum projection methods 
have been developed for zero-temperature cases;
\textit{i.e.} for the ground-state wave functions~\cite{HI79,SGF84},
and also for the quantum-number-constrained HFB (CHFB) solutions 
along the yrast line~\cite{KO81,HHR82,ETY99}. 
For finite-temperature problems, 
projections with respect to the discrete symmetries 
such as the parity and the number-parity 
were explored more than two decades ago~\cite{TSM81}. 
However, there is a certain complication 
in extending them to the continuous symmetries,
arising from the non-commutability between the projection operator 
and the Boltzmann-Gibbs operator in the mean-field approximation. 
A way to overcome these problems has been found in 
Ref.~\cite{EE93} for the particle number projection 
within the Bardeen-Cooper-Schrieffer (BCS) approximation.
General projection formalism
in the variation-before-projection (VBP) scheme
is developed in Ref.~\cite{RR94},
and is extensively applied in the framework
of the static-path approximation.

In this paper we reformulate the quantum number projection
at finite temperature 
by employing the thermo field dynamics (TFD)~\cite{TU75,UMT82}.
The TFD has been shown to be a powerful tool~\cite{TS90}
to handle the thermal fluctuations within the mean-field theories.
We shall show that this is also true 
for the quantum-number-projected HFB theory at finite temperature.
While the resultant formulae are equivalent to those in Ref.~\cite{RR94}, 
the present formalism is advantageous in the following respects. 
In the TFD, the thermal expectation value of an observable 
is expressed in terms of a vacuum expectation value 
in an enlarged TFD space.
Therefore explicit formulae are derived in a straightforward manner, 
keeping complete parallel to the zero-temperature 
formalism~\cite{TEY99}. We demonstrate it for the simultaneous 
projection both of the particle number and the angular momentum.
This leads to another advantage that 
variational parameters can be identified in an obvious manner 
and handled easily.
While application of the present method to the VBP calculations
is straightforward, further approximation is desired
in order to carry out the variation-after-projection (VAP) calculations.
We shall also discuss an approximation of the entropy,
so as for the Peierls inequality~\cite{Pei38} to hold
which guarantees the variational principle for free energy.
It is expected that this approximation scheme will make
the VAP calculations practical.

In Sec.~\ref{sec:TFD}, application of the thermo field dynamics (TFD) 
to the HFB theory at finite temperature is presented. 
In Sec.~\ref{sec:proj}, we formulate the TFD version of 
the quantum-number projection in the HFB at finite temperature.
As well as general arguments, explicit formulae
for the particle number and angular momentum projection are presented
by taking specific basis-sets.
Formulae in the BCS approximation are also given.
In Sec.~\ref{sec:entropy}, we propose an approximation scheme
for the entropy, which keeps the Peierls inequality.
Finally, the paper is summarized in Sec.~\ref{sec:summary}. 

\section{Finite-temperature HFB theory and thermo field dynamics}
\label{sec:TFD}

\subsection{HFB equation at finite temperature} 

In the variational derivation of
the Hartree-Fock-Bogoliubov (HFB) equation 
at finite temperature~\cite{TS80,TSM81}, 
a set of variational parameters 
$\lbrace U_{k\mu}, V_{k\mu}, U_{k\mu}^\ast, V_{k\mu}^\ast\rbrace$ 
is introduced through the generalized Bogoliubov transformation
(GBT), which relates the original single-particle (s.p.) operators 
$\lbrace c_k, c_k^\dagger\rbrace$ to those 
in the quasiparticle (q.p.) picture 
$\lbrace\alpha_\mu, \alpha_\mu^\dagger\rbrace$,
\begin{equation}
\left(\begin{array}{c}c_k \\ c_k^\dagger \end{array}
\right)=\sum_\mu W_{k\mu}\left(\begin{array}{c}\alpha_\mu \\ 
\alpha_\mu^\dagger \end{array}
\right),\quad W_{k\mu}\equiv\left(\begin{array}{cc} U_{k\mu} & 
V_{k\mu}^\ast \\ 
V_{k\mu} & U_{k\mu}^\ast \end{array}\right).  
\label{eq:one} 
\end{equation} 
The transformation matrix $W$ obeys the unitarity relation
$W^\dagger W=W W^\dagger=1$.
We express the original s.p. state as 
$|i\rangle=c_i^\dagger |\mbox{vac.}\rangle$,
with the vacuum $|\mbox{vac.}\rangle$
satisfying $c_i|\mbox{vac.}\rangle=0$.
The q.p. vacuum $|0\rangle\propto\Pi_\nu \alpha_\nu |\mbox{vac.}\rangle$ 
is annihilated by $\alpha_\mu$, \textit{i.e.} $\alpha_\mu|0\rangle=0$.
Note that $\langle 0|0\rangle=1$.
An additional set of variational parameters 
$\lbrace E_\mu\rbrace$ comes into the problem
through the trial statistical operator
\begin{equation}
{\hat w}_0=\frac{e^{-{\hat H}_0/T}} 
{{\rm Tr}(e^{-{\hat H}_0/T})}\,;\quad 
{\hat H}_0=\sum_\mu E_\mu\alpha_\mu^\dagger \alpha_\mu\,,
\label{eq:thr} 
\end{equation} 
where we put the Boltzmann constant $k_{\rm B}=1$. 
We denote the thermal average of an operator ${\hat O}$ by
$\langle{\hat O}\rangle={\rm Tr}({\hat w}_0{\hat O})$, 
where the trace is taken over 
the grand canonical ensemble.

We consider the Hamiltonian comprised of up to two-body interactions, 
\begin{equation}
{\hat H}=\sum_i\varepsilon_i c_i^\dagger c_i
+\frac{1}{4}\sum_{ijkl}v_{ijkl} 
c_i^\dagger c_j^\dagger c_l c_k,    
\label{eq:nin}
\end{equation}
where two kinds of nucleons are not discriminated 
to avoid unnecessary complication. 
The hermiticity of the Hamiltonian implies
$\varepsilon_i=\varepsilon_i^\ast$ and
$v_{ijkl}=-v_{jikl}=-v_{ijlk}=v_{klij}^\ast$.
In the constrained HFB (CHFB) approximation at finite temperature,
the grand potential is given by
\begin{equation}
\Omega=\langle{\hat H}^\prime\rangle -TS\,;\quad
{\hat H}^\prime\equiv {\hat H}-\lambda_{\rm p}{\hat Z}
-\lambda_{\rm n}{\hat N}-\omega_{\rm rot}{\hat J}_x\,,
\label{eq:fiv}
\end{equation}
where $\lambda_{\rm p}$ ($\lambda_{\rm n}$) and 
$\omega_{\rm rot}$ are Lagrange multipliers, 
which are interpreted as proton (neutron) chemical 
potential and rotational frequency of the system, respectively.
The proton- (neutron-) number operator ${\hat Z}$ (${\hat N}$) 
and the $x$-component of angular momentum operator ${\hat J}_x$
are expressed in terms of the s.p. operators as
\begin{equation}
{\hat Z}=\sum_{i\in {\rm p}}c_i^\dagger c_i,\quad
{\hat N}=\sum_{i\in {\rm n}}c_i^\dagger c_i,\quad
{\hat J}_x=\sum_{ij({\rm all})}\langle i|{\hat J}_x|j\rangle
c_i^\dagger c_j, 
\label{eq:six} 
\end{equation} 
where the sum $\sum_{i\in {\rm p}}$ $(\sum_{i\in {\rm n}})$ 
extends over the proton (neutron) states, and 
$\sum_{ij({\rm all})}$ over both proton and neutron states.
In the CHFB the conservations of proton number, neutron number 
and angular momentum are taken into account in their averages,
\begin{equation}
\langle{\hat Z}\rangle=Z,\quad \langle{\hat N}\rangle
=N,\quad \langle{\hat J}_x\rangle=\sqrt{J(J+1)}\,,
\label{eq:fou}
\end{equation}
which indicate constraints.
In calculations without some of the constraints,
the associated Lagrange multipliers are set to be zero
in Eq.~(\ref{eq:fiv}).

The auxiliary Hamiltonian in Eq.~(\ref{eq:fiv}) is rewritten 
in terms of the q.p. operators by the GBT (\ref{eq:one}), 
\begin{equation}
{\hat H}^\prime=U_0+{\hat H}_{11}+{\hat H}_{20}+{\hat H}_{22}
+{\hat H}_{31}+{\hat H}_{40}. 
\label{eq:ele} 
\end{equation} 
For later convenience, explicit forms in the above expression
are presented in Appendix A. 

Ensemble averages of the bilinear forms in the q.p. operators
are given by 
\begin{subequations}
\begin{eqnarray}
\langle\alpha_\mu^\dagger\alpha_\nu\rangle
&=&f_\mu\delta_{\mu\nu}\,;\quad 
f_\mu=\frac{1}{e^{E_\mu/T}+1}\,,\label{eq:seva} \\
\langle\alpha_\mu\alpha_\nu\rangle
&=&\langle\alpha_\mu^\dagger\alpha_\nu^\dagger\rangle=0\,,  
\label{eq:sevb}
\end{eqnarray}
\end{subequations}
where $f_\mu$ is the occupation number of 
the $\mu$-th q.p. level. 
Thus, the variation of the parameter $\delta E_\mu$ is converted
to $\delta f_\mu=-f_\mu(1-f_\mu)\delta E_\mu/T$.
The approximate entropy $S$ in Eq.~(\ref{eq:fiv}) is 
expressed as 
\begin{equation}
S=-{\rm Tr}({\hat w}_0\ln{\hat w}_0) 
=-\sum_\mu [f_\mu\ln f_\mu +(1-f_\mu)\ln(1-f_\mu)]\,. 
\label{eq:eig}
\end{equation}
The s.p. density $\rho$ and the 
pair tensor $\kappa$ are defined by 
\begin{subequations}
\begin{eqnarray}
\rho_{ij}&=&\langle c_j^\dagger c_i\rangle 
=[V^\ast(1-f)V^{\rm tr}+UfU^\dagger]_{ij}\,,\label{eq:fourta} \\
\kappa_{ij}&=&\langle c_jc_i\rangle 
=[V^\ast(1-f)U^{\rm tr}+UfV^\dagger]_{ij}\,.
\label{fourtb} 
\end{eqnarray} 
\label{eq:fourt} 
\end{subequations}
The ensemble average of ${\hat H}^\prime$ becomes 
\begin{equation}
\langle{\hat H}^\prime\rangle={\rm Tr}_\mathrm{s.p.}
\bigg(\xi\rho+\frac{1}{2}\Gamma\rho+
\frac{1}{2}\Delta\kappa^\dagger\bigg)
=U_0+\sum_\mu (H_{11})_{\mu\mu}f_\mu
+2\sum_{\mu\nu}(H_{22})_{\mu\nu\mu\nu}f_\mu f_\nu\,, 
\label{eq:twe}
\end{equation}
where $\mathrm{Tr}_\mathrm{s.p.}$ stands for the trace
over the s.p. space,
\begin{equation}
\xi_{ij}=(\varepsilon_i-\lambda_\tau)\delta_{ij}
-\omega_{\rm rot}\langle i|{\hat J}_x|j\rangle 
\label{eq:thirt}
\end{equation}
with $\tau=\mathrm{p}$ ($\mathrm{n})$ for proton (neutron),
while the HF potential matrix $\Gamma$
and the pair potential matrix $\Delta$ are defined by 
\begin{subequations}
\begin{eqnarray}
\Gamma_{ij}&=&(\Gamma^\dagger)_{ij}
=\sum_{kl}v_{ikjl}\rho_{lk}\,, \label{eq:eighta} \\
\Delta_{ij}&=&-\Delta_{ji}
=\frac{1}{2}\sum_{kl}v_{ijkl}\kappa_{kl}\,. 
\label{eq:eightb}
\end{eqnarray}
\label{eq:eight}
\end{subequations}
The variational principle
$\delta\Omega=\delta(\langle{\hat H}^\prime\rangle-TS)=0$ 
yields~\cite{TSM81} 
\begin{subequations}
\begin{eqnarray}
(H_{11}^{\rm eff})_{\mu\nu} 
&\equiv&\big[U^\dagger(\xi+\Gamma)U
-V^\dagger(\xi+\Gamma)^\ast V +U^\dagger\Delta V
-V^\dagger\Delta^\ast U\big]_{\mu\nu}
\nonumber \\ 
&=&(H_{11})_{\mu\nu}
+4\sum_\rho(H_{22})_{\mu\rho\nu\rho}f_\rho \nonumber \\
&=&T\frac{\partial S}{\partial f_\mu}\delta_{\mu\nu}
=E_\mu\delta_{\mu\nu}\,, \label{eq:ninta} \\ 
(H_{20}^{\rm eff})_{\mu\nu} 
&\equiv&\big[U^\dagger(\xi+\Gamma)V^\ast
-V^\dagger(\xi+\Gamma)^\ast U^\ast +U^\dagger\Delta U^\ast
-V^\dagger\Delta^\ast V^\ast\big]_{\mu\nu}
\nonumber \\
&=&(H_{20})_{\mu\nu}
+6\sum_\rho(H_{31})_{\mu\nu\rho\rho}f_\rho =0\,. 
\label{eq:nintb}
\end{eqnarray}
\label{eq:nint}
\end{subequations}
These equations are summarized in the form of 
the HFB coupled equation at finite temperature, 
\begin{equation}
\sum_l\left(\begin{array}{cc} \xi+\Gamma & \Delta \\
-\Delta^\ast & -(\xi+\Gamma)^\ast \end{array}\right)_{kl}
\left(\begin{array}{c} U \\ V \end{array}\right)_{l\mu}
=\left(\begin{array}{c} U \\ V \end{array}\right)_{k\mu}E_\mu\,. 
\label{eq:twent}
\end{equation}
In the CHFB case this equation should be solved
together with the constraints in Eq.~(\ref{eq:fou}),
by which the Lagrange multipliers $\lambda_{\rm p}$,
$\lambda_{\rm n}$ and $\omega_{\rm rot}$,
as well as self-consistent q.p. energies $E_\mu$,
are determined as functions of quantum numbers and temperature. 

\subsection{Bogoliubov transformation extended by TFD} 

In the thermo field dynamics (TFD)~\cite{TU75,UMT82},
the original q.p. operator space 
$\lbrace \alpha_\mu, \alpha_\mu^\dagger\rbrace$ 
is enlarged by including newly introduced tilded operators;
$\lbrace \alpha_\mu, {\tilde \alpha}_\mu, 
\alpha_\mu^\dagger, {\tilde \alpha}_\mu^\dagger\rbrace$.
Correspondingly, the q.p. vacuum in the enlarged space
is defined by $|0\rangle\otimes|\tilde{0}\rangle$,
where $|\tilde{0}\rangle$ is the tilded vacuum
satisfying $\tilde{\alpha}_\mu|\tilde{0}\rangle=0$.
The TFD vacuum $|0\rangle\otimes|\tilde{0}\rangle$
is hereafter denoted by $|0\rangle$ for the sake of simplicity,
\textit{i.e.} $\alpha_\mu|0\rangle=\tilde{\alpha}_\mu|0\rangle=0$.
In order to handle ensemble averages
in a thermal equilibrium at temperature $T$,
the TFD vacuum $|0\rangle$ is transformed
to the temperature-dependent vacuum
by the following unitary transformation,
\begin{equation}
|0_T\rangle=e^{-\hat{G}}|0\rangle,\quad \hat{G}=-\hat{G}^\dagger
=\sum_\mu\vartheta_\mu(\alpha_\mu^\dagger{\tilde \alpha}_\mu^\dagger
-{\tilde \alpha}_\mu \alpha_\mu),
\label{eq:twtwo}
\end{equation}
where 
\begin{equation}
\sin\vartheta_\mu=f_\mu^{1/2}\equiv g_\mu,\quad 
\cos \vartheta_\mu=(1-f_\mu)^{1/2}\equiv {\bar g}_\mu. 
\label{eq:twthr}
\end{equation}
This unitary transformation relates the operator set 
$\lbrace \alpha_\mu, {\tilde \alpha}_\mu, 
\alpha_\mu^\dagger, {\tilde \alpha}_\mu^\dagger\rbrace$ 
to new set of the temperature-dependent 
operators $\lbrace \beta_\mu,{\tilde \beta}_\mu, 
\beta_\mu^\dagger, {\tilde \beta}_\mu^\dagger\rbrace$, 
\begin{eqnarray} 
\alpha_\mu&=&e^{\hat{G}}\beta_\mu e^{-\hat{G}}={\bar g}_\mu\beta_\mu 
+g_\mu{\tilde \beta}_\mu^\dagger, \nonumber \\ 
{\tilde \alpha}_\mu&=&e^{\hat{G}}{\tilde \beta}_\mu e^{-\hat{G}} 
={\bar g}_\mu{\tilde \beta}_\mu -g_\mu \beta_\mu^\dagger, 
\label{eq:twfou} 
\end{eqnarray} 
so that the temperature-dependent vacuum should fulfill
$\beta_\mu|0_T\rangle={\tilde \beta}_\mu|0_T\rangle=0$.
The temperature-dependent quasiparticles
created by $\beta^\dagger$ or $\tilde{\beta}^\dagger$
will hereafter be called TFD quasiparticles. 
The thermal average of an observable ${\hat O}$
is then expressed by the vacuum expectation value at $|0_T\rangle$,
\begin{equation}
\langle{\hat O}\rangle={\rm Tr}({\hat w}_0{\hat O})
=\langle 0_T|{\hat O}|0_T\rangle. 
\label{eq:twone} 
\end{equation} 

Now we unify two unitary transformations (\ref{eq:one}) 
and (\ref{eq:twfou}) to compose an extended form of the GBT~\cite{TS90}, 
\begin{equation}
\left(\begin{array}{c} c \\ {\tilde c} \\ c^\dagger \\ 
{\tilde c}^\dagger \end{array}\right)_k=\sum_\mu{\bar W}_{k\mu}
\left(\begin{array}{c} \beta \\ {\tilde \beta} \\ 
\beta^\dagger \\ {\tilde \beta}^\dagger 
\end{array}\right)_\mu,\quad
{\bar W}_{k\mu}=\left(\begin{array}{cc} 
{\bar U} & {\bar V}^\ast \\ {\bar V} & {\bar U}^\ast
\end{array}\right)_{k\mu},
\label{eq:twsix}
\end{equation}
where 
\begin{equation}
{\bar U}_{k\mu}=\left(\begin{array}{cc} U{\bar g} & V^\ast g \\ 
-V^\ast g & U{\bar g} \end{array}\right)_{k\mu},\quad
{\bar V}_{k\mu}=\left(\begin{array}{cc} V{\bar g} & U^\ast g \\ 
-U^\ast g & V{\bar g} \end{array}\right)_{k\mu}.
\label{eq:twsev}
\end{equation}
Because of the unitarity relation ${\bar W}{\bar W}^\dagger=
{\bar W}^\dagger{\bar W}=1$, the matrix inverse to ${\bar W}$ is 
given by 
\begin{equation}
{\bar W}^{-1}={\bar W}^\dagger=\left(\begin{array}{cc} 
{\bar U}^\dagger & {\bar V}^\dagger \\
{\bar V}^{\rm tr} & {\bar U}^{\rm tr}\end{array}\right). 
\label{eq:tweig}
\end{equation} 

\section{QUANTUM NUMBER PROJECTION in finite-temperature HFB theory}
\label{sec:proj}

\subsection{Projection operators}

\subsubsection{General form}

Since quantum numbers are usually associated
with a certain group structure of the system,
projection with respect to the quantum numbers is also introduced
in connection to the group.
In general, a projection operator 
${\hat P}_{\mu\nu}^{(\alpha)}$ is defined as a 
subring basis corresponding to an irreducible representation (irrep.)
$\varrho^{(\alpha)}$ of a group ${\cal G}$~\cite{Ham64,Nor80}, 
\begin{equation}
{\hat P}_{\mu\nu}^{(\alpha)}=
\frac{{\rm dim}(\varrho^{(\alpha)})}{g}\sum_{x\in{\cal G}}
\varrho_{\nu\mu}^{(\alpha)}(x^{-1})\,x\,, 
\label{eq:twnin}
\end{equation} 
where $x$ stands for an element of ${\cal G}$,
${\rm dim}(\varrho^{(\alpha)})$ is the dimension of the 
representation matrix $\varrho^{(\alpha)}$,
and $g$ the order of ${\cal G}$.
The property of group representation yields 
\begin{equation}
{\hat P}_{\mu\nu}^{(\alpha)}
{\hat P}_{\mu^\prime\nu^\prime}^{(\beta)}=
\delta_{\alpha\beta}\delta_{\nu\mu^\prime}
{\hat P}_{\mu\nu^\prime}^{(\alpha)}\,,
\label{eq:thirty}
\end{equation}
justifying a convenient bracket representation of 
the projection operator, ${\hat P}_{\mu\nu}^{(\alpha)}
=|\alpha\mu\rangle\langle\alpha\nu|$.
The projection on the subspace corresponding
to the irrep. $\varrho^{(\alpha)}$,
without referencing $\mu$, is obtained by
\begin{equation}
{\hat P}_\alpha = \sum_\mu {\hat P}_{\mu\mu}^{(\alpha)}
= \sum_\mu |\alpha\mu\rangle\langle\alpha\mu|.
\label{eq:thirtyb}
\end{equation}
It is obvious that $\hat{P}_\alpha$ is idempotent,
\textit{i.e.} $\hat{P}_\alpha^2=\hat{P}_\alpha$.

When $\mathcal{G}$ is not simple
and is decomposed into a direct product of an invariant subgroup
and the complementary quotient group as
$\mathcal{G}_1 \otimes \mathcal{G}_2$,
its irrep. $\varrho^{(\alpha)}$ is also a product
such as $\varrho_1^{(\alpha_1)}\otimes\varrho_2^{(\alpha_2)}$,
where $\varrho_i^{(\alpha_i)}$ is an irrep. of $\mathcal{G}_i$
($i=1,2$).
Denoting the projection operator on $\varrho_i^{(\alpha_i)}$
by $\hat{P}_{\alpha_i}$,
the projection operator on $\varrho^{(\alpha)}$
is $\hat{P}_\alpha = \hat{P}_{\alpha_1}\hat{P}_{\alpha_2}$.
Therefore, projection operators of simple groups
are essential.

In the following we assume $\mathcal{G}$
to be a compact simple group or a product of such groups,
whose element $x$ is represented by an appropriate unitary operator
$\hat{Q}=e^{-i\hat{S}}$.
Here $\hat{S}=\hat{S}(\Theta)$ is a hermitian operator,
which belongs to a Lie algebra associated with $\mathcal{G}$
and is dependent on a set of parameters $\Theta$.
The projection operator $\hat{P}_\alpha$ is represented
in the integral form
\begin{equation}
{\hat P}_\alpha = \int d\Theta\,\zeta_\alpha(\Theta)\,\hat{Q},
\label{eq:thirtyc}
\end{equation}
where $\zeta_\alpha(\Theta)$ is an appropriate function of $\Theta$
and is derived from Eq.~(\ref{eq:twnin}).

\subsubsection{Angular momentum projection operator}

For the angular momentum projection, the relevant group 
is $SU(2)$, which is denoted by $SU(2)_J$ in this paper.
The group element is the rotation operator ${\hat R}$,
whose parameters are represented by $\Phi$.
In practice, we consider two alternative parameterizations.
One is the Euler angles $\Phi=(\alpha, \beta, \gamma)$.
The rotation is also represented by an angle $\omega$
$(0\leq \omega<2\pi)$ around an axis
indicated by a unit vector ${\vec n}$.
Using the representation
$n_x=\sin\theta\cos\phi, n_y=\sin\theta\sin\phi, n_z=\cos\theta$,
we arrive at the other parameterization $\Phi=(\omega,\theta,\phi)$.
Corresponding to these parameterizations,
the rotation operator is expressed in two ways as 
\begin{equation}
{\hat R}(\Phi)
=e^{-i\alpha{\hat J}_z}e^{-i\beta{\hat J}_y}e^{-i\gamma{\hat J}_z}
=e^{-i\omega(n_x{\hat J}_x+n_y{\hat J}_y+n_z{\hat J}_z)}\,, 
\label{eq:thone}
\end{equation} 
where the angular momentum operators are defined in terms of 
the s.p. operators as in Eq.~(\ref{eq:six}),
\begin{equation}
{\hat J}_k=\sum_{ij({\rm all})}\langle i|{\hat J}_k|j\rangle 
c_i^\dagger c_j \qquad 
(k=x, y, z). 
\label{eq:thtwo}
\end{equation}
The angles $(\omega, \theta, \phi)$ are related to the Euler angles by
\begin{equation}
\cos\frac{\omega}{2}=\cos\frac{\beta}{2}
\cos\frac{\alpha+\gamma}{2}\,,\quad
\tan\theta=\tan\frac{\beta}{2}\bigg/\sin\frac{\alpha+\gamma}{2}\,,\quad
\phi=\frac{\pi+\alpha-\gamma}{2}\,.
\label{eq:ththr}
\end{equation}
The order of the group is determined by the volume of 
the parameter space, 
\begin{equation}
g=\int_0^{2\pi}d\alpha\int_0^\pi\sin\beta d\beta\int_0^{4\pi}
d\gamma=4\int_0^{2\pi}\sin^2\frac{\omega}{2}d\omega\int_0^\pi
\sin\theta d\theta
\int_0^{2\pi}d\phi=16\pi^2\,, 
\label{eq:thfou}
\end{equation}
where the range of $\gamma$ is taken to be 
$0\leq \gamma< 4\pi$ for the double-valued 
representation corresponding to half-odd-integer spin. 
The irrep. of $SU(2)_J$ is given
by the Wigner $D$-function~\cite{Ros57,Var88},
\begin{equation}
D_{MK}^J(\Phi)\equiv\langle JM|{\hat R}(\Phi)|JK\rangle 
=e^{-i\alpha M}d_{MK}^J(\beta) e^{-i\gamma K}\,,
\label{eq:thfiv}
\end{equation}
which is a unitary matrix of dimension $(2J+1)$.
Then Eq.~(\ref{eq:twnin}) yields the following projection operator,
\begin{equation}
{\hat P}^J_{MK}={\hat P}_{MK}^{J\dagger}
=\frac{2J+1}{16\pi^2}\int_0^{2\pi}d\alpha\int_0^\pi
\sin\beta d\beta \int_0^{4\pi}d\gamma\,
D_{MK}^{J^\ast}(\Phi) {\hat R}(\Phi). \label{eq:thsix}
\end{equation}
For integer spin $J$, the integral 
$\int_0^{4\pi}d\gamma$ may 
be replaced by $2\int_0^{2\pi}d\gamma$. 

We express the character of the representation matrix 
$D_{MK}^J(\Phi)$ as 
\begin{equation}
\chi_J(\omega)=\sum_{M=-J}^J D_{MM}^J(\Phi)
=\frac{\sin[(J+1/2)\omega]}{\sin(\omega/2)}. 
\label{eq:thsev}
\end{equation}
The real function $\chi_J(\omega)$ can be represented in many 
different ways as listed in Ref.~\cite{Var88}.
If the magnitude of angular momentum $J$ is subject
to the projection
while its $z$-component in the laboratory frame $M$ is not referenced, 
the projection operator is given by 
\begin{equation}
{\hat P}_J={\hat P}_J^\dagger\equiv
\sum_{M=-J}^J{\hat P}_{MM}^J
=\frac{2J+1}{4\pi^2}\int_0^{2\pi}\sin^2\frac{\omega}{2}d\omega
\int_0^\pi\sin\theta d\theta\int_0^{2\pi}d\phi\,
\chi_J(\omega) {\hat R}(\Phi)\,.
\label{eq:theig} 
\end{equation} 

\subsubsection{Particle number projection operators} 

The relevant group to the particle number projection
is the gauge group $U(1)$,
whose group elements are parameterized by a single variable $\varphi$
for the particle number operator ${\hat N}$;
$\exp(-i\varphi {\hat N})$.
For neutrons, the group is denoted by $U(1)_N$,
and the projection operator, which projects out states having 
the exact neutron number $N$, is given by, 
\begin{equation}
{\hat P}_N={\hat P}_N^\dagger=\frac{1}{2\pi}\int_0^{2\pi}
e^{-i\varphi_{\rm n}({\hat N}-N)}d\varphi_{\rm n}.
\label{eq:thninb}
\end{equation}
Likewise, the proton number projection is implemented
by the operator,
\begin{equation}
{\hat P}_Z={\hat P}_Z^\dagger=\frac{1}{2\pi}\int_0^{2\pi}
e^{-i\varphi_{\rm p}({\hat Z}-Z)}d\varphi_{\rm p},
\label{eq:thnina}
\end{equation}
and the relevant group to $\hat{P}_Z$ is denoted by $U(1)_Z$.
The product operator $\hat{P}_Z\hat{P}_N$ is employed
for the $U(1)_Z\times U(1)_N$ projection,
by which the canonical trace is calculable
from the grand-canonical trace.

\subsubsection{$SU(2)_J\times U(1)_Z\times U(1)_N$ projection operator} 

The projector of simultaneous projection of 
angular momentum $J$, proton number $Z$ and neutron number $N$ is 
${\hat P}_{(J,Z,N)}={\hat P}_J{\hat P}_Z{\hat P}_N$,
which satisfies ${\hat P}_{(J,Z,N)}^2={\hat P}_{(J,Z,N)}$.
This $\hat{P}_{(J,Z,N)}$ can be expressed
in the form of Eq.~(\ref{eq:thirtyc}),
with $\Theta=(\Phi,\varphi_\mathrm{p},\varphi_\mathrm{n})$.
The operator ${\hat Q}$ stands for 
\begin{equation}
{\hat Q}={\hat R}(\Phi)\,e^{-i\varphi_{\rm p} {\hat Z}}
e^{-i\varphi_{\rm n} {\hat N}}= e^{-i{\hat S}},
\label{eq:fifty}
\end{equation} 
where
\begin{equation}
{\hat S}=\varphi_{\rm p}{\hat Z}+\varphi_{\rm n}{\hat N}
+\omega\,{\vec n}\cdot{\vec {\hat J}}; \quad 
{\vec n}\cdot{\vec {\hat J}}\equiv 
n_x{\hat J}_x+n_y{\hat J}_y+n_z{\hat J}_z\,.
\label{eq:fifone}
\end{equation}
Obviously ${\hat P}_{(J,Z,N)}$
commutes with the total nuclear Hamiltonian ${\hat H}$, 
which conserves angular momentum, nucleon number and charge. 
However, note that $[{\hat P}_{(J,Z,N)}, {\hat H}_0]\not=0$~\cite{RR94}. 

\subsubsection{Parity and number-parity projection operators}

The space reflection forms the discrete group isomorphic to $S_2$.
The projection operator with respect to the space parity
can be expressed in the form similar to Eq.~(\ref{eq:thirtyc}),
with the integral converted to a discrete sum,
\begin{equation}
{\hat P}_\pi=\frac{1}{2}\bigg[1+\xi 
\exp\,\big(i\pi\sum_k^- c_k^\dagger c_k\big)\bigg]\,.
\label{eq:fotone}
\end{equation}
Here $\xi=+1$ $(\xi=-1)$ for the projection of positive (negative) 
parity state, and $\sum_k^-$ denotes the sum extending only 
over the s.p. states of negative parity. 
As far as the parity is not mixed in the q.p. state $\mu$,
the projection operator is also written as
\begin{equation}
{\hat P}_\pi=\frac{1}{2}\bigg[1+\xi 
\exp\,\big(i\pi\sum_\mu^-\alpha_\mu^\dagger\alpha_\mu\big)\bigg]\,,
\label{eq:fotonea}
\end{equation}
as given in Ref.~\cite{TSM81}.

The number-parity is relevant to another $S_2$ group,
which is a discrete subgroup of $U(1)_Z$ or of $U(1)_N$.
Respective to protons and neutrons,
the number-parity projection is carried out by the operator
\begin{equation} 
{\hat P}_q=\frac{1}{2}\bigg[1+\eta 
\exp\,\big(i\pi\sum_k c_k^\dagger c_k\big)\bigg]\,,
\label{eq:fottwo2} 
\end{equation} 
where $\eta=+1$ $(\eta=-1)$ for the projection of the state of 
even (odd) number-parity.
Apart from the denominator, this is obtained
by restricting $\varphi_\mathrm{n}$ in $\hat{P}_N$
(or $\varphi_\mathrm{p}$ in $\hat{P}_Z$) only to $0$ and $\pi$.
Since the GBT (\ref{eq:one}) does not mix the number-parity,
$\hat{P}_q$ is equivalently represented
in terms of the q.p. operators~\cite{TSM81},
\begin{equation} 
{\hat P}_q=\frac{1}{2}\bigg[1+\eta 
\exp\,\big(i\pi\sum_\mu \alpha_\mu^\dagger\alpha_\mu\big)\bigg]\,.
\label{eq:fottwo} 
\end{equation} 

\subsection{Extended form of linear transformation}

We now consider the projected statistics.
The projected ensemble average of an operator ${\hat O}$
is newly defined by
\begin{equation}
\langle{\hat O}\rangle_{\rm P}\equiv
{\rm Tr}({\hat w}_{\rm P}{\hat O})
=\frac{{\rm Tr}({\hat P}e^{-\beta{\hat H}_0}{\hat P}
{\hat O}{\hat P})}
{{\rm Tr}({\hat P}e^{-\beta{\hat H}_0}{\hat P})} 
=\frac{{\rm Tr}(e^{-\beta{\hat H}_0}{\hat P}{\hat O}{\hat P})}
{{\rm Tr}(e^{-\beta{\hat H}_0}{\hat P})},
\label{eq:fotthr}
\end{equation}
where the statistical operator with projection 
is introduced as 
\begin{equation}
{\hat w}_{\rm P}\equiv 
\frac{{\hat P}e^{-\beta{\hat H}_0}{\hat P}}
{{\rm Tr}(e^{-\beta{\hat H}_0}{\hat P})}\,. 
\label{eq:fotfou}
\end{equation}
For an operator commutable with $\hat{P}$,
$\langle{\hat O}\rangle_{\rm P}$ is calculated in the TFD by
\begin{eqnarray}
\langle{\hat O}\rangle_{\rm P}
=\frac{{\rm Tr}({\hat w}_0{\hat O}{\hat P})}
{{\rm Tr}({\hat w}_0{\hat P})}
=\frac{\langle 0_T|{\hat O}{\hat P}|0_T\rangle}
{\langle 0_T|{\hat P}| 0_T\rangle}.
\label{eq:fotthr2}
\end{eqnarray}
Substituting the projector by the form of Eq.~(\ref{eq:thirtyc}),
we obtain
\begin{eqnarray}
\langle{\hat O}\rangle_{\rm P}
={\displaystyle\frac{\int \zeta(\Theta)\,
\langle 0_T|{\hat O}{\hat Q}|0_T\rangle\,d\Theta}
{\int \zeta(\Theta)\,\langle 0_T|{\hat Q}| 0_T\rangle\,d\Theta}}.
\label{eq:fotthr3}
\end{eqnarray}
Thus, in course of projection calculations, 
we meet with the TFD vacuum expectation values of following types:
\begin{equation}
\langle 0_T|\left\lbrace\begin{array}{c} 1 \\ 
\beta_1\beta_2 \\ \beta_1\beta_2\beta_3\beta_4 
\end{array}\right\rbrace {\hat Q}|0_T\rangle, 
\label{eq:fotnin}
\end{equation} 
where $\beta$ represents any of the TFD q.p. operator of four types,
$\beta_\mu$, $\beta_\mu^\dagger$, ${\tilde \beta}_\mu$ 
or ${\tilde \beta}_\mu^\dagger$.
The operator ${\hat Q}$ is given in
Eqs.~(\ref{eq:fifty},\ref{eq:fifone})
for the $SU(2)_J\times U(1)_Z\times U(1)_N$ projection.

We here define the transformation matrix of the s.p. operator
with respect to the unitary transformation $\hat{Q}$,
assuming that $\hat{Q}$ conserves the particle number,
\begin{equation}
\hat{Q}c_k\hat{Q}^\dagger = \sum_{k^\prime}Q_{kk^\prime}\,c_{k^\prime}.
\label{eq:fiftwo}
\end{equation}
This linear transformation is extended to the TFD space by
\begin{equation}
{\hat Q}\left(\begin{array}{c} c_k \\ {\tilde c}_k \\
c_k^\dagger \\ {\tilde c}_k^\dagger \end{array}\right){\hat Q}^\dagger
=\sum_{k^\prime}\left(\begin{array}{cccc}
Q_{kk^\prime} & 0 & 0 & 0 \\ 0 & 1 & 0 & 0 \\
0 & 0 & Q_{kk^\prime}^\ast & 0 \\ 0 & 0 & 0 & 1 \end{array}\right)
\left(\begin{array}{c}
c_{k^\prime} \\ {\tilde c}_{k^\prime} \\ 
c_{k^\prime}^\dagger \\ {\tilde c}_{k^\prime}^\dagger
\end{array}\right). 
\label{eq:fifthr}
\end{equation}
Applying the TFD-extension of the GBT
in Eqs.~(\ref{eq:twsix},\ref{eq:twsev},\ref{eq:tweig}), we convert 
the above transformation to the one among the TFD q.p. operators,
\begin{eqnarray}
{\hat Q}\left(\begin{array}{c} \beta_\mu \\ {\tilde \beta}_\mu \\ 
\beta_\mu^\dagger \\ {\tilde \beta}_\mu^\dagger 
\end{array}\right){\hat Q}^\dagger 
&=&\sum_{k,k^\prime}\sum_\nu
({\bar W}^\dagger)_{\mu k}\left(\begin{array}{cccc}
Q_{kk^\prime} & 0 & 0 & 0 \\ 0 & 1 & 0 & 0 \\
0 & 0 & Q_{kk^\prime}^\ast & 0 \\ 0 & 0 & 0 & 1 \end{array}\right)
({\bar W})_{k^\prime \nu}
\left(\begin{array}{c}
\beta_\nu \\ {\tilde \beta}_\nu \\ \beta_\nu^\dagger \\ 
{\tilde \beta}_\nu^\dagger\end{array}\right) \nonumber \\
&\equiv& \sum_\nu\left(\begin{array}{cc} {\bar K} & {\bar N} \\ 
{\bar M} & {\bar L} \end{array}\right)_{\mu\nu}
\left(\begin{array}{c}
\beta_\nu \\ {\tilde \beta}_\nu \\ \beta_\nu^\dagger \\ 
{\tilde \beta}_\nu^\dagger\end{array}\right). 
\label{eq:fiffou}
\end{eqnarray}
In the last expression, there appear four types of the matrices 
${\bar K}$, ${\bar L}$, ${\bar M}$ and ${\bar N}$, 
which play central roles in the projection 
calculation~\cite{HI79,TEY99}.
In the present formalism,
these matrices, whose dimension is also doubled, are given by 
\begin{subequations}
\begin{eqnarray}
{\bar K}_{\mu\nu}&=&{\bar L}_{\mu\nu}^\ast 
=\left(\begin{array}{cc}
{\bar g}(U^\dagger Q U + V^\dagger Q^\ast V){\bar g}+g^2 & 
{\bar g}(U^\dagger Q V^\ast + V^\dagger Q^\ast U^\ast)g \\
g(V^{\rm tr} Q U + U^{\rm tr} Q^\ast V){\bar g} & 
g(V^{\rm tr} Q V^\ast + U^{\rm tr} Q^\ast U^\ast)g+{\bar g}^2 
\end{array}\right)_{\mu\nu}, \label{eq:fiffiva} \\
{\bar M}_{\mu\nu}&=&{\bar N}_{\mu\nu}^\ast 
=\left(\begin{array}{cc}
{\bar g}(V^{\rm tr} Q U + U^{\rm tr} Q^\ast V){\bar g} & 
{\bar g}(V^{\rm tr} Q V^\ast + U^{\rm tr} Q^\ast U^\ast)g - g{\bar g} \\ 
g(U^\dagger Q U + V^\dagger Q^\ast V){\bar g} - {\bar g}g &  
g(U^\dagger Q V^\ast + V^\dagger Q^\ast U^\ast)g
\end{array}\right)_{\mu\nu}, 
\label{eq:fiffivb}
\end{eqnarray}
\label{eq:fiffiv}
\end{subequations}
where we have employed abbreviated notations, \textit{e.g.},
\begin{equation}
[{\bar g}(U^\dagger Q U + V^\dagger Q^\ast V)g+g^2]_{\mu\nu}
=\sum_{k,k^\prime}{\bar g}_\mu(U_{k\mu}^\ast Q_{kk^\prime} U_{k^\prime\nu}
+V_{k\mu}^\ast {Q_{kk^\prime}}^\ast V_{k^\prime\mu})g_\nu
+g_\mu^2\delta_{\mu\nu}.
\label{eq:fifsix}
\end{equation}

\subsection{Wick's theorem and reduction formulae}

Corresponding to Eq.~(\ref{eq:fiftwo}),
the operator ${\hat S}$ in $\hat{Q}=e^{-i\hat{S}}$ has the form
\begin{equation}
{\hat S}=\sum_{k,k^\prime} S_{kk^\prime} c_k^\dagger c_{k^\prime}\,.
\label{eq:sevtwoa}
\end{equation}
According to the general prescription
of projection calculation~\cite{TEY99},
we represent ${\hat S}$ in terms of the TFD q.p. operators as 
\begin{equation}
{\hat S}= \bar{S}^{(0)} + \sum_{\mu,\nu ({\rm all})}\bar{S}_{\mu\nu}^{(1)}
\beta_\mu^\dagger\beta_\nu
+\frac{1}{2}\sum_{\mu,\nu ({\rm all})}
(\bar{S}_{\mu\nu}^{(2)}\beta_\mu^\dagger\beta_\nu^\dagger + \mbox{h.c.}), 
\label{eq:sevtwo}
\end{equation}
where the summation $\sum_{\mu,\nu ({\rm all})}$ extends over both 
the tilded and non-tilded q.p. states.
On the other hand, the TFD-extension of the GBT derives
an alternative form as
\begin{eqnarray}
{\hat S}&=&\sum_{\mu,\,\nu >0}\left(\begin{array}{cccc}
\beta_\mu^\dagger & \tilde{\beta}_\mu^\dagger &
\beta_\mu & \tilde{\beta}_\mu \end{array}\right)
\left(\begin{array}{cc} \bar{S}_{11} & \bar{S}_{12} \\
 \bar{S}_{21} & \bar{S}_{22} \end{array}\right)_{\mu\nu}
\left(\begin{array}{c} \beta_\nu \\ \tilde{\beta}_\nu \\
\beta_\nu^\dagger \\ \tilde{\beta}_\nu^\dagger 
\end{array}\right) \nonumber \\
&=&{\rm Tr}(\bar{S}_{22})+\sum_{\mu,\nu}\left\{
\left(\begin{array}{cc}
\beta_\mu^\dagger & \tilde{\beta}_\mu^\dagger \end{array}\right)
(\bar{S}_{11}-\bar{S}_{22}^{\rm tr})_{\mu\nu}
\left(\begin{array}{c} \beta_\nu \\ \tilde{\beta}_\nu
\end{array}\right)
+ \Big[\left(\begin{array}{cc}
\beta_\mu^\dagger & \tilde{\beta}_\mu^\dagger \end{array}\right)
(\bar{S}_{12})_{\mu\nu}\left(\begin{array}{c}
\beta_\nu^\dagger \\ \tilde{\beta}_\nu^\dagger 
\end{array}\right)
+ \mbox{h.c.} \Big]\right\}, \nonumber\\
\label{eq:sevthr}
\end{eqnarray}
where $\bar{S}_{11}$, $\bar{S}_{12}$, $\bar{S}_{21}$ and $\bar{S}_{22}$
are calculated as 
\begin{subequations}
\begin{eqnarray}
(\bar{S}_{11})_{\mu\nu}&=&\left(\begin{array}{cc}
({\bar g}U^\dagger SU{\bar g})_{\mu\nu} & 
({\bar g}U^\dagger SV^\ast g)_{\mu\nu} \\
(gV^{\rm tr}SU{\bar g})_{\mu\nu} & 
(gV^{\rm tr}SV^\ast g)_{\mu\nu} 
\end{array}\right), \label{eq:sevfoua} \\
(\bar{S}_{22})_{\mu\nu}&=&\left(\begin{array}{cc} 
({\bar g}V^{\rm tr}SV^\ast{\bar g})_{\mu\nu} & 
({\bar g}V^{\rm tr}SUg)_{\mu\nu} \\
(gU^\dagger SV^\ast{\bar g})_{\mu\nu} & 
(gU^\dagger SUg)_{\mu\nu}
\end{array}\right), \label{eq:sevfoub} \\
(\bar{S}_{12})_{\mu\nu}&=&(\bar{S}_{21}^\dagger)_{\mu\nu}=
\left(\begin{array}{cc}
({\bar g}U^\dagger SV^\ast{\bar g})_{\mu\nu} & 
({\bar g}U^\dagger SUg)_{\mu\nu} \\
(gV^{\rm tr}SV^\ast{\bar g})_{\mu\nu} & 
(gV^{\rm tr}SUg)_{\mu\nu}
       \end{array}\right). \label{eq:sevfouc}
\end{eqnarray}
\label{eq:sevfou}
\end{subequations}
Comparing (\ref{eq:sevtwo}) with the last expression 
in (\ref{eq:sevthr}), 
we obtain 
\begin{equation}
\bar{S}^{(0)}={\rm Tr}\bar{S}_{22}\,, \quad 
\bar{S}^{(1)}=\bar{S}_{11}-\bar{S}_{22}^{\rm tr}\,, \quad 
\bar{S}^{(2)}=2\bar{S}_{12}=2\bar{S}_{21}^\dagger\,.
\label{eq:sevfiv}
\end{equation}
By applying the general formalism of projection~\cite{TEY99},
the TFD vacuum expectation value of the operator ${\hat Q}$ is given by 
\begin{equation}
\langle 0_T|{\hat Q}|0_T\rangle 
=({\rm det} {\bar K})^{1/2}\exp\!\left[-i\bigg(\bar{S}^{(0)}
+\frac{1}{2}{\rm Tr}\bar{S}^{(1)}\bigg)\right]
=({\rm det} {\bar K})^{1/2}\exp\!\left[-\frac{i}{2}{\rm Tr}S\right].  
\label{eq:sevsix}
\end{equation}
It is noted that $\mathrm{Tr}S$ vanishes for the traceless algebra
like that associated with $SU(2)_J$,
while gives dimension of the s.p. space for $U(1)_Z$ and $U(1)_N$.

In order to apply the generalized Wick's theorem to 
reduce the quantities in (\ref{eq:fotnin}), 
we introduce an operator symbol $[{\hat Q}]$~\cite{TEY99}, 
and express the TFD vacuum expectation value of 
an operator ${\hat O}$ as  
\begin{equation}
\langle 0_T|{\hat O}{\hat Q}|0_T\rangle
=\langle 0_T|{\hat Q}|0_T\rangle 
\langle 0_T|{\hat O}[{\hat Q}]|0_T\rangle\,;\quad 
[{\hat Q}]\equiv\frac{{\hat Q}}
{\langle 0_T|{\hat Q}|0_T\rangle}. 
\label{eq:eighty}
\end{equation}
Then, the basic matrix elements generalized by the TFD,
which are building blocks for the quantum number projection
at finite temperature, are provided by 
\begin{subequations} 
\begin{eqnarray} 
A_{\mu\nu}\equiv\langle 0_T|[{\hat Q}]
\beta_\mu^\dagger\beta_\nu^\dagger|0_T\rangle 
&=&[{\bar M}{\bar K}^{-1}]_{\mu\nu} 
=-[({\bar K}^{\rm tr})^{-1}{\bar M}^{\rm tr}]_{\mu\nu}, 
\label{eq:eigonea} \\ 
B_{\mu\nu}\equiv\langle 0_T|
\beta_\mu\beta_\nu[{\hat Q}]|0_T\rangle 
&=&[{\bar K}^{-1}{\bar N}]_{\mu\nu}
=-[{\bar N}^{\rm tr}({\bar K}^{\rm tr})^{-1}]_{\mu\nu}, 
\label{eq:eigoneb} \\ 
C_{\mu\nu}\equiv\langle 0_T|\beta_\mu[{\hat Q}]
\beta_\nu^\dagger|0_T\rangle 
&=&[{\bar K}^{-1}]_{\mu\nu}, \label{eq:eigonec}
\end{eqnarray} 
\label{eq:eigone} 
\end{subequations} 
where the suffices $\mu$ and $\nu$ represent
all the possible q.p. states.
The TFD version of the generalized Wick's theorem~\cite{TEY99} is
exemplified by
\begin{subequations}
\begin{eqnarray}
\langle 0_T|[{\hat Q}]\beta_\mu^\dagger\beta_\nu^\dagger
\beta_\rho^\dagger\beta_\sigma^\dagger|0_T\rangle
&=&A_{\mu\nu}A_{\rho\sigma}-A_{\mu\rho}A_{\nu\sigma}
+A_{\mu\sigma}A_{\nu\rho}, \label{eq:eigtwoa} \\
\langle 0_T|\beta_\mu[{\hat Q}]\beta_\nu^\dagger
\beta_\rho^\dagger\beta_\sigma^\dagger|0_T\rangle
&=&C_{\mu\nu}A_{\rho\sigma}-C_{\mu\rho}A_{\nu\sigma}
+C_{\mu\sigma}A_{\nu\rho}, \label{eq:eigtwob} \\
\langle 0_T|\beta_\mu\beta_\nu[{\hat Q}]
\beta_\rho^\dagger\beta_\sigma^\dagger|0_T\rangle
&=&B_{\mu\nu}A_{\rho\sigma}-C_{\mu\rho}C_{\nu\sigma}
+C_{\mu\sigma}C_{\nu\rho}, \label{eq:eigtwoc} \\
\langle 0_T|\beta_\mu\beta_\nu
\beta_\rho[{\hat Q}]\beta_\sigma^\dagger|0_T\rangle
&=&B_{\mu\nu}C_{\rho\sigma}-B_{\mu\rho}C_{\nu\sigma}
+C_{\mu\sigma}B_{\nu\rho}, \label{eq:eigtwod} \\
\langle 0_T|\beta_\mu\beta_\nu
\beta_\rho\beta_\sigma[{\hat Q}]|0_T\rangle
&=&B_{\mu\nu}B_{\rho\sigma}-B_{\mu\rho}B_{\nu\sigma}
+B_{\mu\sigma}B_{\nu\rho}. \label{eq:eigtwoe} 
\end{eqnarray}
\label{eq:eigtwo}
\end{subequations}

For the $SU(2)_J\times U(1)_Z\times U(1)_N$ projection,
we have 
\begin{eqnarray}
\frac{1}{2}{\rm Tr}S=\varphi_{\rm p}\Omega_{\rm p}
+\varphi_{\rm n}\Omega_{\rm n},  
\label{eq:sevsev}
\end{eqnarray}
where the quantity $\Omega_{\rm p}=\sum_{j\in{\rm p}}(j+1/2)$ 
($\Omega_{\rm n}=\sum_{j\in{\rm n}}(j+1/2)$) represents 
a half-number of total s.p. levels in the model space.
Hence, we obtain 
\begin{eqnarray}
\langle 0_T|{\hat P}_{(J,Z,N)}|0_T\rangle&=&\frac{1}{2\pi}\int_0^{2\pi}
e^{-i\varphi_{\rm p}(\Omega_{\rm p}-Z)}d\varphi_{\rm p}\,
\frac{1}{2\pi}\int_0^{2\pi}
e^{-i\varphi_{\rm n}(\Omega_{\rm n}-N)}d\varphi_{\rm n} 
\nonumber \\
&&\quad\times \frac{2J+1}{4\pi^2}\int_0^{2\pi}
 \sin^2\frac{\omega}{2}d\omega
\int_0^\pi\sin\theta d\theta\int_0^{2\pi}d\phi\,
\chi_J(\omega)({\rm det}{\bar K})^{1/2},
\label{eq:sevnin}
\end{eqnarray}
and
\begin{eqnarray}
\langle 0_T|{\hat H}{\hat P}_{(J,Z,N)}|0_T\rangle
&=&\frac{1}{2\pi}\int_0^{2\pi}
e^{-i\varphi_{\rm p}(\Omega_{\rm p}-Z)}d\varphi_{\rm p}\,
\frac{1}{2\pi}\int_0^{2\pi}
e^{-i\varphi_{\rm n}(\Omega_{\rm n}-N)}d\varphi_{\rm n}
\nonumber \\
&&\times
\frac{2J+1}{4\pi^2}\int_0^{2\pi}\sin^2\frac{\omega}{2}d\omega
\int_0^\pi\sin\theta d\theta\int_0^{2\pi}d\phi\,
\chi_J(\omega)({\rm det}{\bar K})^{1/2}
\langle 0_T|{\hat H}[{\hat Q}]|0_T\rangle. \nonumber\\
\label{eq:eigthr}
\end{eqnarray}
Applying Eq.~(\ref{eq:eigtwo}) to
$\langle 0_T|{\hat H}[{\hat Q}]|0_T\rangle$,
we can compute the Hamiltonian kernel in Eq.~(\ref{eq:eigthr}).
The $SU(2)_J\times U(1)_Z\times U(1)_N$ projected
thermal energy is calculated from Eq.~(\ref{eq:fotthr2}),
by carrying out integrals in Eqs.~(\ref{eq:sevnin},\ref{eq:eigthr}).
It is pointed out that the method of 
integration, which has been demonstrated to be successful 
in the zero-temperature case~\cite{HI79,ETY99},
is available also for the projected statistics. If we choose 
Euler angles $(\alpha, \beta, \gamma)$ rather than 
the variables $(\omega, \theta, \phi)$ for 
integral variables, the Gauss-Chebyshev quadrature 
formula will be useful for the $\alpha$ and 
$\gamma$ integrals as well as the $\varphi_{\rm p}$ and 
$\varphi_{\rm n}$ integrals, and the Gauss-Legendre 
quadrature formula for the $\beta$ integral, since 
the function $\chi_J(\omega)$ is nothing but a sum of 
the Wigner $D$-functions.

It should be noticed that the original q.p. occupation number 
is affected by the projection. Using the TFD 
transformation in (\ref{eq:twfou}), we get 
\begin{equation}
\alpha_\mu^\dagger\alpha_\mu=
f_\mu+(1-f_\mu)\beta_\mu^\dagger\beta_\mu
-\sqrt{f_\mu(1-f_\mu)}\beta_\mu{\tilde \beta}_\mu
+\sqrt{f_\mu(1-f_\mu)}\beta_\mu^\dagger{\tilde \beta}_\mu^\dagger. 
\label{eq:ninety}
\end{equation}
The first and third terms in the above expression
contribute to the projected occupation number,
when the projection operator is placed on the right of 
$\alpha_\mu^\dagger\alpha_\mu$.
For $\hat{P}_{(J,Z,N)}$,
the projected q.p. occupation number is calculated by 
\begin{eqnarray}
\langle 0_T|\alpha_\mu^\dagger\alpha_\mu {\hat P}_{(J,Z,N)}
|0_T\rangle &=&\frac{1}{2\pi}\int_0^{2\pi}
e^{-i\varphi_{\rm p}(\Omega_{\rm p}-Z)}d\varphi_{\rm p}\,
\frac{1}{2\pi}\int_0^{2\pi}
e^{-i\varphi_{\rm n}(\Omega_{\rm n}-N)}d\varphi_{\rm n} \nonumber\\
&&\qquad\times \frac{2J+1}{4\pi^2}\int_0^{2\pi}\sin^2
\frac{\omega}{2}d\omega
\int_0^\pi\sin\theta d\theta\int_0^{2\pi}d\phi \nonumber \\
&&\qquad\times\chi_J(\omega)
({\rm det}{\bar K})^{1/2}\big[f_\mu-\sqrt{f_\mu(1-f_\mu)}
({\bar K}^{-1}{\bar N})_{\mu\tilde{\mu}}\big].
\label{eq:ninone}
\end{eqnarray}

The present formalism is summarized by
Eqs.~(\ref{eq:fotthr3},\ref{eq:sevsix},\ref{eq:eighty},\ref{eq:eigtwo})
for general cases,
with $A_{\mu\nu}$, $B_{\mu\nu}$, $C_{\mu\nu}$ of Eq.~(\ref{eq:eigone}),
and $\bar{K}_{\mu\nu}=\bar{L}_{\mu\nu}^\ast$,
$\bar{M}_{\mu\nu}=\bar{N}_{\mu\nu}^\ast$ of Eq.~(\ref{eq:fiffiv}),
which are functions of the GBT coefficients $\{U_{k\mu}, V_{k\mu},
U_{k\mu}^\ast, V_{k\mu}^\ast\}$ and the q.p. occupation numbers $f_\mu$.
The integral in Eq.~(\ref{eq:fotthr3}) is exemplified
by Eqs.~(\ref{eq:sevnin},\ref{eq:eigthr},\ref{eq:ninone})
for the $SU(2)_J\times U(1)_Z\times U(1)_N$ projection.
The resultant formulae are equivalent
to those derived in Ref.~\cite{RR94},
despite the difference in appearance.
It still deserves noting that
the present formulae are straightforward extension
of the zero-temperature formulae of Refs.~\cite{ETY99,TEY99}.

\subsection{Representation in specific bases}

For practical use, we here present explicit forms of the quantities
relevant to the $SU(2)_J\times U(1)_Z\times U(1)_N$ projection,
taking specific s.p. bases.

\subsubsection{Spherical bases}

Each s.p. level $|\tau nljm\rangle$ is specified by 
$\tau(={\rm p},{\rm n})$, radial quantum number $n$,
orbital angular momentum $l$,
total angular momentum $j$ $(=l\pm 1/2)$ 
and its projection to the $z$-axis $m$. 
Here the s.p. state $k$ is regarded
as an abbreviation of $(\tau nljm)$, namely,
$|k\rangle\equiv |\tau nljm\rangle$.
Analogously, $|k^\prime\rangle\equiv
|\tau^\prime n^\prime l^\prime j^\prime m^\prime\rangle$.
For the projection ${\hat P}_{(J,Z,N)}$,
we have $\hat{Q}=e^{-i\varphi_\tau}{\hat R}$,
yielding the $Q$-matrix in Eq.~(\ref{eq:fiftwo}) of
\begin{equation}
Q_{kk^\prime}
=Q_{\tau nljm, \tau^\prime n^\prime l^\prime j^\prime m^\prime}
= \delta_{\tau\tau^\prime}\delta_{nn^\prime}\delta_{ll^\prime}
\delta_{jj^\prime}\,e^{i\varphi_\tau} D_{m^\prime m}^{j^\ast}.
\label{eq:fiftwo2}
\end{equation}
The ${\bar K}$, ${\bar L}$, ${\bar M}$ and ${\bar N}$ matrices
are explicitly written as 
\begin{subequations}
\begin{eqnarray}
{\bar K}_{\mu\nu}&=&{\bar L}_{\mu\nu}^\ast 
=\left(\begin{array}{cc}
{\bar g}(U^\dagger e^{i\varphi_\tau}D^\dagger U
+V^\dagger e^{-i\varphi_\tau}D^{\rm tr}V){\bar g}+g^2 & 
{\bar g}(U^\dagger e^{i\varphi_\tau}D^\dagger V^\ast
+V^\dagger e^{-i\varphi_\tau}D^{\rm tr}U^\ast)g \\
g(V^{\rm tr}e^{i\varphi_\tau}D^\dagger U
+U^{\rm tr}e^{-i\varphi_\tau}D^{\rm tr}V){\bar g} & 
g(V^{\rm tr}e^{i\varphi_\tau}D^\dagger V^\ast 
+U^{\rm tr} e^{-i\varphi_\tau}D^{\rm tr}U^\ast)g+{\bar g}^2 
\end{array}\right)_{\mu\nu}, \nonumber\\
\label{eq:fiffiva2} \\
{\bar M}_{\mu\nu}&=&{\bar N}_{\mu\nu}^\ast 
=\left(\begin{array}{cc}
{\bar g}(V^{\rm tr}e^{i\varphi_\tau}D^\dagger U
+U^{\rm tr}e^{-i\varphi_\tau}D^{\rm tr}V){\bar g} & 
{\bar g}(V^{\rm tr}e^{i\varphi_\tau}D^\dagger V^\ast
+U^{\rm tr}e^{i\varphi_\tau}D^\dagger U^\ast)g
-g{\bar g} \\ 
g(U^\dagger e^{i\varphi_\tau}D^\dagger U
+V^\dagger e^{-i\varphi_\tau}D^{\rm tr}V){\bar g}
-{\bar g}g &  
g(U^\dagger e^{i\varphi_\tau}D^\dagger V^\ast
+V^\dagger e^{-i\varphi_\tau}D^{\rm tr}U^\ast)g
\end{array}\right)_{\mu\nu}, \nonumber\\
\label{eq:fiffivb2}
\end{eqnarray}
\label{eq:fiffiv2}
\end{subequations}
where, \textit{e.g.},
\begin{eqnarray}
&&[{\bar g}(U^\dagger e^{i\varphi_\tau}D^\dagger U
+V^\dagger e^{-i\varphi_\tau}D^{\rm tr}V)g
+g^2]_{\mu\nu} \nonumber\\
&&\quad =\sum_{k,k^\prime}
\delta_{\tau\tau^\prime}\delta_{nn^\prime}\delta_{ll^\prime}
\delta_{jj^\prime}\,{\bar g}_\mu(U_{k\mu}^\ast e^{i\varphi_\tau}
D_{m^\prime m}^{j\ast}U_{k^\prime\nu}+V_{k\mu}^\ast 
e^{-i\varphi_\tau}D_{m^\prime m}^j
V_{k^\prime\mu})g_\nu+g_\mu^2\delta_{\mu\nu}.
\label{eq:fifsix2}
\end{eqnarray}
When we take only the particle number projection, 
all $D_{m^\prime m}^j$ should be replaced by $\delta_{mm^\prime}$.

\subsubsection{Goodman's bases}

Whole spherical basis states for a given set of quantum numbers 
$(\tau nlj)$ can be separated into two sets 
$\lbrace |\tau nljm\rangle\rbrace$ 
and $\lbrace |\tau\overline{nljm}\rangle\rbrace$. The first set 
$\lbrace|\tau nljm\rangle\rbrace$ consists of the states having 
$(m-1/2)$ equal to even integers, while the second set 
$\lbrace|\tau\overline{nljm}\rangle\equiv {\hat T}|\tau nljm\rangle
\rbrace$ having $(m-1/2)$ equal to odd integers, where ${\hat T}$ is 
the time-reversal transformation.
The phase convention is assumed to be 
\begin{equation}
|\tau\overline{nljm}\rangle\equiv {\hat T}|\tau nljm\rangle
=(-1)^{j-m+l}|\tau nlj-m\rangle, \quad 
{\hat T}^2|\tau nljm\rangle=-|\tau nljm\rangle. 
\label{eq:fifsev}
\end{equation}
Since the auxiliary Hamiltonian in Eq.~(\ref{eq:fiv}) is invariant 
under the rotation about the $x$-axis by angle $\pi$, 
\textit{i.e.} ${\hat R}_x=e^{-i\pi{\hat J}_x}$, 
the solution of the CHFB equation usually
becomes eigenstates of ${\hat R}_x$.
Therefore, a convenient choice is to employ the set of 
eigenstates of $\hat{R}_x$ for s.p. basis. 
From the relations as 
\begin{equation}
{\hat R}_xc_{\tau nljm}^\dagger {\hat R}_x^\dagger
=e^{-i\pi j}c_{\tau nlj-m}^\dagger, \quad 
({\hat R}_x)^2c_{\tau nljm}^\dagger({\hat R}_x^\dagger)^2
=-c_{\tau nljm}^\dagger, 
\label{eq:fifeig}
\end{equation}
we find that the eigenvalues of ${\hat R}_x$ are $\pm i$, 
\begin{equation}
{\hat R}_xc_k^\dagger{\hat R}_x^\dagger=-i c_k^\dagger, \quad 
{\hat R}_xc_{\bar k}^\dagger{\hat R}_x^\dagger
=+i c_{\bar k}^\dagger\,,
\label{eq:fifnin}
\end{equation}
where the corresponding eigenoperators are given by 
\begin{equation}
c_k^\dagger=\frac{1}{\sqrt{2}}
\big[c_{\tau nljm}^\dagger+(-1)^{j-1/2}c_{\tau nlj-m}^\dagger\big], 
\quad 
c_{\bar k}^\dagger=\frac{(-1)^{l+m+1/2}}{\sqrt{2}}
\big[c_{\tau nljm}^\dagger-(-1)^{j-1/2}c_{\tau nlj-m}^\dagger\big].
\label{eq:sixty}
\end{equation}
The operator $\hat{R}_x$ is called signature operator,
and the bases $(k\bar{k})$ are called Goodman's bases~\cite{Goo74}.
It is sufficient to adopt the bases of Eq.~(\ref{eq:sixty})
only for even $(m-1/2)$ to avoid repetition,
which will symbolically be indicated by $k>0$.
Using the relations in Eq.~(\ref{eq:fifsev}), we can confirm that 
$c_k^\dagger$ and $c_{\bar k}^\dagger$ are 
related to each other by the time-reversal transformation, 
\begin{equation}
{\hat T}c_k^\dagger{\hat T}^{-1}=c_{\bar k}^\dagger, \quad 
{\hat T}c_{\bar k}^\dagger{\hat T}^{-1}=-c_k^\dagger.
\label{eq:sixone}
\end{equation} 
Accordingly matrix forms for expectation values of 
relevant quantities are also simplified. As an example, 
we here present the angular momentum components, 
\begin{eqnarray}
({\hat J}_x)_{(k\bar{k}),(k^\prime\bar{k^\prime})}&=&
\left(\begin{array}{cc} \langle k|{\hat J}_x|k^\prime\rangle & 0 \\
0 & \langle {\bar k}|{\hat J}_x|{\bar k}^\prime\rangle\end{array}\right),
\quad 
({\hat J}_y)_{(k\bar{k}),(k^\prime\bar{k^\prime})}=
\left(\begin{array}{cc} 0 & \langle k|{\hat J}_y|{\bar k^\prime}\rangle \\
\langle {\bar k}|{\hat J}_y|k^\prime\rangle & 0 \end{array}\right),
\nonumber \\
({\hat J}_z)_{(k\bar{k}),(k^\prime\bar{k^\prime})}&=&
\left(\begin{array}{cc} 0 & \langle k|{\hat J}_z|{\bar k}^\prime\rangle \\
\langle {\bar k}|{\hat J}_z|k^\prime\rangle & 0 \end{array}\right).
\label{eq:sixthr}
\end{eqnarray}
Due to time-reversal properties, we have the following relations,
\begin{equation}
\langle k|{\hat J}_x|k^\prime\rangle=
 -\langle {\bar k^\prime}|{\hat J}_x|{\bar k}\rangle,\quad 
\langle k|{\hat J}_y|{\bar k^\prime}\rangle=
 \langle k^\prime|{\hat J}_y|{\bar k}\rangle,\quad 
\langle k|{\hat J}_z|{\bar k^\prime}\rangle=
 \langle k^\prime|{\hat J}_z|{\bar k}\rangle\,.
\label{eq:sixfou}
\end{equation}

Unless the signature is spontaneously violated,
the q.p. operators $\alpha_\mu^\dagger$ and $\alpha_{\bar \mu}^\dagger$
belonging to the HFB eigenvalues $E_\mu$ and $E_{\bar \mu}$,
respectively, are classified in a similar manner, 
\begin{equation}
{\hat R}_x\alpha_\mu^\dagger {\hat R}_x^\dagger
 =-i\alpha_\mu^\dagger,\quad 
{\hat R}_x\alpha_{\bar \mu}^\dagger {\hat R}_x^\dagger
 =+i\alpha_{\bar \mu}^\dagger\quad\mbox{and}\quad 
{\hat T}\alpha_\mu^\dagger{\hat T}^{-1}
=\alpha_{\bar \mu}^\dagger, \quad 
{\hat T}\alpha_{\bar \mu}^\dagger{\hat T}^{-1}
=-\alpha_\mu^\dagger.
\label{eq:sixtwo}
\end{equation} 
The conservation of the signature simplifies the structure 
of the GBT matrix as 
\begin{equation}
\left(\begin{array}{c} c_k \\ c_{\bar k} \\ 
c_k^\dagger \\ c_{\bar k}^\dagger \end{array}\right)
=\sum_{\mu>0}\left(\begin{array}{cccc} U_{k\mu} & 0 & 
 0 & V_{k{\bar \mu}}^\ast \\
 0 & U_{{\bar k}{\bar \mu}} & V_{{\bar k}\mu}^\ast & 0 \\
 0 & V_{k{\bar \mu}} & U_{k\mu}^\ast & 0 \\
 V_{{\bar k}\mu} & 0 & 0 & U_{{\bar k}{\bar \mu}}^\ast 
\end{array}\right)\left(\begin{array}{c} \alpha_\mu \\ 
\alpha_{\bar \mu} \\ \alpha_\mu^\dagger \\ 
\alpha_{\bar \mu}^\dagger \end{array}\right). 
\label{eq:sixfiv}
\end{equation}
Therefore, the TFD-extension of the GBT is represented as 
\begin{equation}
\left(\begin{array}{c} {\cal C}_{(k{\bar k})} \\
{\cal C}_{(k{\bar k})}^\dagger\end{array}\right)=
\sum_{\mu >0}
 {\bar W}_{(k{\bar k}),\,(\mu{\bar \mu})}
\left(\begin{array}{c}{\cal B}_{(\mu{\bar \mu})} \\
{\cal B}_{(\mu{\bar \mu})}^\dagger\end{array}\right);\quad 
{\cal C}_{(k{\bar k})}\equiv 
\left(\begin{array}{cc} c_k \\ c_{\bar k} \\ 
{\tilde c}_k \\ {\tilde c}_{\bar k}
\end{array}\right),\quad 
{\cal B}_{(\mu{\bar \mu})}\equiv 
\left(\begin{array}{cc} \beta_\mu \\ \beta_{\bar \mu} \\ 
{\tilde \beta}_\mu \\ {\tilde \beta}_{\bar \mu} 
\end{array}\right).
\label{eq:sixsix}
\end{equation}
The transformation matrix $\bar{W}$ is given by
\begin{subequations}
\begin{eqnarray}
{\bar W}_{(k{\bar k}),\,(\mu{\bar \mu})}
&=&\left(\begin{array}{cc}
{\bar U}_{(k{\bar k}),\,(\mu{\bar \mu})} & 
{\bar V}_{(k{\bar k}),\,(\mu{\bar \mu})}^\ast \\
{\bar V}_{(k{\bar k}),\,(\mu{\bar \mu})} & 
{\bar U}_{(k{\bar k}),\,(\mu{\bar \mu})}^\ast 
\end{array}\right); \; \\
{\bar U}_{(k{\bar k}),\,(\mu{\bar \mu})}
&\equiv&\left(\begin{array}{cccc} 
U_{k\mu}{\bar g}_\mu & 0 & 0 & V_{k{\bar \mu}}^\ast g_{\bar \mu} \\
0 & U_{{\bar k}{\bar \mu}}{\bar g}_{\bar \mu} 
& V_{{\bar k}\mu}^\ast g_\mu & 0 \\ 
0 & -V_{k{\bar \mu}}^\ast g_{\bar \mu} & 
U_{k\mu}{\bar g}_\mu & 0 \\
-V_{{\bar k}\mu}^\ast g_\mu & 0 & 0 & 
U_{{\bar k}{\bar \mu}}{\bar g}_{\bar \mu}
\end{array}\right),\nonumber \\
{\bar V}_{(k{\bar k}),\,(\mu{\bar \mu})}
&\equiv&\left(\begin{array}{cccc} 
0 & V_{k{\bar \mu}}{\bar g}_{\bar \mu} & U_{k\mu}^\ast g_\mu & 0 \\
V_{{\bar k}\mu}{\bar g}_\mu & 0 & 0 & 
U_{{\bar k}{\bar \mu}}^\ast g_{\bar \mu} \\ 
-U_{k\mu}^\ast g_\mu & 0 & 0 & 
V_{k{\bar \mu}}{\bar g}_{\bar \mu} \\
0 & -U_{{\bar k}{\bar \mu}}^\ast g_{\bar \mu} & 
V_{{\bar k}\mu}{\bar g}_\mu & 0
\end{array}\right). 
\end{eqnarray}
\label{eq:sixsev}
\end{subequations}

We now consider the $SU(2)_J\times U(1)_Z\times U(1)_N$ projection.
The operator $\hat{Q}=e^{-i\varphi_\tau}{\hat R}$ gives
the transformation rules for the s.p. operators~\cite{ETY99},
\begin{equation}
\hat{Q}c_k\hat{Q}^\dagger=e^{i\varphi_\tau}\sum_{k^\prime >0}
\bigg(F_{kk^\prime}^\ast c_{k^\prime}
+ G_{k\bar{k^\prime}}^\ast c_{\bar{k^\prime}}\bigg), \quad
\hat{Q}c_{\bar k}\hat{Q}^\dagger=e^{i\varphi_\tau}\sum_{k^\prime >0}
\bigg(G_{{\bar k}k^\prime}c_{k^\prime}
+F_{{\bar k}\bar{k^\prime}}c_{\bar{k^\prime}}\bigg),
\label{eq:sixeig}
\end{equation}
where the coefficients $F$ and $G$ are defined by 
\begin{subequations}
\begin{eqnarray}
F_{kk^\prime}&=&F_{{\bar k}\bar{k^\prime}} 
=\delta_{\tau\tau^\prime}\delta_{nn^\prime}\delta_{ll^\prime}
\delta_{jj^\prime}\,
\frac{1}{2}\bigg[D_{m^\prime, m}^j+D_{-m^\prime, -m}^j 
+(-1)^{j-1/2}(D_{m^\prime, -m}^j+D_{-m^\prime, m}^j)\bigg], \nonumber\\
\label{eq:sixnina} \\
G_{k\bar{k^\prime}}&=&-G_{{\bar k}k^\prime} \nonumber\\
&=&\delta_{\tau\tau^\prime}\delta_{nn^\prime}\delta_{ll^\prime}
\delta_{jj^\prime}\,
\frac{(-1)^{l+m^\prime+1/2}}{2}
\bigg[D_{m^\prime, m}^j-D_{-m^\prime, -m}^j
+(-1)^{j-1/2}(D_{m^\prime, -m}^j-D_{-m^\prime, m}^j)\bigg].
\nonumber \\
\label{eq:sixninb}
\end{eqnarray}
\label{eq:sixnin}
\end{subequations}
The transformation (\ref{eq:sixeig}) is 
converted to the one among the TFD q.p. operators
via Eqs.~(\ref{eq:sixsix},\ref{eq:sixsev}), 
\begin{equation}
{\hat Q}\left(\begin{array}{c} {\cal B}_{(\mu{\bar \mu})} \\ 
{\cal B}_{(\mu{\bar \mu})}^\dagger \end{array}\right){\hat Q}^\dagger 
=\sum_{\nu >0}
\left(\begin{array}{cc} {\bar K}_{(\mu{\bar \mu}),\,(\nu{\bar \nu})} & 
{\bar N}_{(\mu{\bar \mu}),\,(\nu{\bar \nu})} \\ 
{\bar M}_{(\mu{\bar \mu}),\,(\nu{\bar \nu})} & 
{\bar L}_{(\mu{\bar \mu}),\,(\nu{\bar \nu})} \end{array}\right) 
\left(\begin{array}{c} {\cal B}_{(\nu{\bar \nu})} \\ 
{\cal B}_{(\nu{\bar \nu)}}^\dagger \end{array}\right), 
\label{eq:seventy}
\end{equation}
where the submatrices ${\bar K}$, ${\bar L}$, ${\bar M}$ 
and ${\bar N}$ are provided by
\begin{subequations}
\begin{eqnarray}
&&{\bar K}_{(\mu\,{\bar \mu}),\,(\nu{\bar \nu})}
={\bar L}_{(\mu{\bar \mu}),\,(\nu{\bar \nu})}^\ast \nonumber \\
&&\quad=\left(\begin{array}{cc} 
\bar{g}_\mu(U^\dagger e^{i\varphi_\tau}F^\ast U
+V^\dagger e^{-i\varphi_\tau}F^\ast V)_{\mu\nu}\bar{g}_\nu & 
\bar{g}_\mu(U^\dagger e^{i\varphi_\tau}G^\ast U
+V^\dagger e^{-i\varphi_\tau}G^\ast V)_{\mu{\bar \nu}}\bar{g}_{\bar{\nu}} \\
\bar{g}_{\bar{\mu}}(U^\dagger e^{i\varphi_\tau}GU
+V^\dagger e^{-i\varphi_\tau}GV)_{{\bar \mu}\nu}\bar{g}_\nu & 
\bar{g}_{\bar{\mu}}(U^\dagger e^{i\varphi_\tau}FU
+V^\dagger e^{-i\varphi_\tau}FV)_{{\bar \mu}{\bar \nu}}\bar{g}_{\bar{\nu}} \\ 
g_\mu(V^{\rm tr}e^{i\varphi_\tau}GU
+U^{\rm tr}e^{-i\varphi_\tau}GV)_{\mu\nu}\bar{g}_\nu & 
g_\mu(V^{\rm tr}e^{i\varphi_\tau}FU
+U^{\rm tr}e^{-i\varphi_\tau}FV)_{\mu{\bar \nu}}\bar{g}_{\bar{\nu}}  \\
g_{\bar{\mu}}(V^{\rm tr}e^{i\varphi_\tau}F^\ast U
+U^{\rm tr}e^{-i\varphi_\tau}F^\ast V)_{{\bar \mu}\nu}\bar{g}_\nu & 
g_{\bar{\mu}}(V^{\rm tr}e^{i\varphi_\tau}G^\ast U
+U^{\rm tr}e^{-i\varphi_\tau}G^\ast V)_{{\bar \mu}{\bar \nu}}
\bar{g}_{\bar{\nu}} 
\end{array}\right. \nonumber \\
&&\qquad~\left.\begin{array}{cc}
\bar{g}_\mu(U^\dagger e^{i\varphi_\tau}G^\ast V^\ast
+V^\dagger e^{-i\varphi_\tau}G^\ast U^\ast)_{\mu\nu}g_\nu &
\bar{g}_\mu(U^\dagger e^{i\varphi_\tau}F^\ast V^\ast
+U^\dagger e^{-i\varphi_\tau}F^\ast V^\ast)_{\mu{\bar \nu}}g_{\bar{\nu}} \\
\bar{g}_{\bar{\mu}}(U^\dagger e^{i\varphi_\tau}FV^\ast
+V^\dagger e^{-i\varphi_\tau}FU^\ast)_{{\bar \mu}\nu}g_\nu &
\bar{g}_{\bar{\mu}}(U^\dagger e^{i\varphi_\tau}GV^\ast
+V^\dagger e^{-i\varphi_\tau}GU^\ast)_{{\bar \mu}{\bar \nu}}g_{\bar{\nu}} \\
g_\mu(V^{\rm tr}e^{i\varphi_\tau}FV^\ast
+U^{\rm tr}e^{-i\varphi_\tau}FU^\ast)_{\mu\nu}g_\nu &
g_\mu(V^{\rm tr}e^{i\varphi_\tau}GV^\ast
+U^{\rm tr}e^{-i\varphi_\tau}GU^\ast)_{\mu{\bar \nu}}g_{\bar{\nu}} \\
g_{\bar{\mu}}(V^{\rm tr}e^{i\varphi_\tau}G^\ast V^\ast
+U^{\rm tr}e^{-i\varphi_\tau}G^\ast U^\ast)_{{\bar \mu}\nu}g_\nu &
g_{\bar{\mu}}(V^{\rm tr}e^{i\varphi_\tau}F^\ast V^\ast
+U^{\rm tr}e^{-i\varphi_\tau}F^\ast U^\ast)_{{\bar \mu}{\bar \nu}}g_{\bar{\nu}}
\end{array}\right) \nonumber\\
&&\qquad +~\delta_{\mu\nu} \left(\begin{array}{cccc} 
g_\mu^2 & 0 & 0 & 0 \\ 0 & g_{\bar{\mu}}^2 & 0 & 0 \\
0 & 0 & \bar{g}_\mu^2 & 0 \\ 0 & 0 & 0 & \bar{g}_{\bar{\mu}}^2
\end{array}\right), \label{eq:sevonea} \\
&&{\bar M}_{(\mu{\bar \mu}),\,(\nu{\bar \nu})}
={\bar N}_{(\mu{\bar \mu}),\,(\nu{\bar \nu})}^\ast \nonumber \\ 
&&\quad=\left(\begin{array}{cc} 
\bar{g}_\mu(V^{\rm tr}e^{i\varphi_\tau}GU
+U^{\rm tr}e^{-i\varphi_\tau}GV)_{\mu\nu}\bar{g}_\nu & 
\bar{g}_\mu(V^{\rm tr}e^{i\varphi_\tau}FU
+U^{\rm tr}e^{-i\varphi_\tau}FV)_{\mu{\bar \nu}}\bar{g}_{\bar{\nu}} \\ 
\bar{g}_{\bar{\mu}}(V^{\rm tr}e^{i\varphi_\tau}F^\ast U
+U^{\rm tr}e^{-i\varphi_\tau}F^\ast V)_{{\bar \mu}\nu}\bar{g}_\nu & 
\bar{g}_{\bar{\mu}}(V^{\rm tr}e^{i\varphi_\tau}G^\ast U
+U^{\rm tr}e^{-i\varphi_\tau}G^\ast V)_{{\bar \mu}{\bar \nu}}
\bar{g}_{\bar{\nu}}  \\ 
g_\mu(U^\dagger e^{i\varphi_\tau}F^\ast U
+V^\dagger e^{-i\varphi_\tau}F^\ast V)_{\mu\nu}\bar{g}_\nu & 
g_\mu(U^\dagger e^{i\varphi_\tau}G^\ast U
+V^\dagger e^{-i\varphi_\tau}G^\ast V)_{\mu{\bar \nu}}\bar{g}_{\bar{\nu}}  \\
g_{\bar{\mu}}(U^\dagger e^{i\varphi_\tau}GU
+V^\dagger e^{-i\varphi_\tau}GV)_{{\bar \mu}\nu}\bar{g}_\nu & 
g_{\bar{\mu}}(U^\dagger e^{i\varphi_\tau}FU
+V^\dagger e^{-i\varphi_\tau}FV)_{{\bar \mu}{\bar \nu}}\bar{g}_{\bar{\nu}} 
\end{array}\right. \nonumber \\
&&\qquad~\left.\begin{array}{cc}
\bar{g}_\mu(V^{\rm tr}e^{i\varphi_\tau}FV^\ast
+U^{\rm tr}e^{-i\varphi_\tau}FU^\ast)_{\mu\nu}g_\nu &
\bar{g}_\mu(V^{\rm tr}e^{i\varphi_\tau}GV^\ast
+U^{\rm tr}e^{-i\varphi_\tau}GU^\ast)_{\mu{\bar \nu}}g_{\bar{\nu}} \\
\bar{g}_{\bar{\mu}}(V^{\rm tr}e^{i\varphi_\tau}G^\ast V^\ast
+U^{\rm tr}e^{-i\varphi_\tau}G^\ast U^\ast)_{{\bar \mu}\nu}g_\nu &
\bar{g}_{\bar{\mu}}(V^{\rm tr}e^{i\varphi_\tau}F^\ast V^\ast
+U^{\rm tr}e^{-i\varphi_\tau}F^\ast U^\ast)_{{\bar \mu}{\bar \nu}}
g_{\bar{\nu}} \\
g_\mu(U^\dagger e^{i\varphi_\tau}G^\ast V^\ast
+V^\dagger e^{-i\varphi_\tau}G^\ast U^\ast)_{\mu\nu}g_\nu &
g_\mu(U^\dagger e^{i\varphi_\tau}F^\ast V^\ast
+V^\dagger e^{-i\varphi_\tau}F^\ast U^\ast)_{\mu{\bar \nu}}g_{\bar{\nu}} \\
g_{\bar{\mu}}(U^\dagger e^{i\varphi_\tau}FV^\ast
+V^\dagger e^{-i\varphi_\tau}FU^\ast)_{{\bar \mu}\nu}g_\nu &
g_{\bar{\mu}}(U^\dagger e^{i\varphi_\tau}GV^\ast
+V^\dagger e^{-i\varphi_\tau}GU^\ast)_{{\bar \mu}{\bar \nu}}g_{\bar{\nu}}
\end{array}\right)\nonumber\\
&&\qquad -~\delta_{\mu\nu} \left(\begin{array}{cccc} 
0 & 0 & g_\mu \bar{g}_\mu & 0 \\ 
0 & 0 & 0 & g_{\bar{\mu}}\bar{g}_{\bar{\mu}} \\ 
\bar{g}_\mu g_\mu & 0 & 0 & 0 \\ 
0 & \bar{g}_{\bar{\mu}}g_{\bar{\mu}} & 0 & 0
\end{array}\right).
\label{eq:sevoneb}
\end{eqnarray}
\label{eq:sevone}
\end{subequations}
Note that, for the case of particle number projection only, 
all the matrices $F$ should be replace by an identity, 
and $G$ by zero.

\subsection{Formulae in BCS approximation}

We continue to work with the Goodman's bases.
The HFB approximation is reduced to the BCS approximation
by the replacements 
\begin{equation}
U_{k\mu}=U_{{\bar k}{\bar \mu}}=u_k\delta_{k\mu},\quad 
V_{k{\bar \mu}}=-V_{{\bar k}\mu}=v_k\delta_{k\mu}. 
\label{eq:eigfou}
\end{equation}
We here assume $u_k$ and $v_k$ to be real.
Thereby the BCS transformation is expressed as 
\begin{equation}
\left(\begin{array}{c} c_k \\ c_{\bar k} \\ c_k^\dagger \\ 
c_{\bar k}^\dagger\end{array}\right)=
\left(\begin{array}{cccc}
u_k & 0 & 0 & v_k \\ 0 & u_k & -v_k & 0 \\
0 & v_k & u_k & 0 \\ -v_k & 0 & 0 & u_k 
\end{array}\right)
\left(\begin{array}{c}
\alpha_k \\ \alpha_{\bar k} \\ \alpha_k^\dagger \\ 
\alpha_{\bar k}^\dagger
\end{array}\right). 
\label{eq:eigfiv}
\end{equation}
Accordingly, the quantum number projection is somewhat simplified
in the BCS approximation.
By using the notations 
\begin{eqnarray}
\xi_{kk^\prime}&=&u_ku_{k^\prime}e^{i\varphi_\tau}
 +v_kv_{k^\prime}e^{-i\varphi_\tau}
=(u_ku_{k^\prime}+v_kv_{k^\prime})\cos\varphi_\tau 
+i(u_ku_{k^\prime}-v_kv_{k^\prime})\sin\varphi_\tau, \nonumber \\
\eta_{kk^\prime}&=&u_kv_{k^\prime}e^{i\varphi_\tau}
 -v_ku_{k^\prime}e^{-i\varphi_\tau}
=(u_kv_{k^\prime}-v_ku_{k^\prime})\cos\varphi_\tau
+i(u_kv_{k^\prime}+v_ku_{k^\prime})\sin\varphi_\tau,
\label{eq:eigsix}
\end{eqnarray}
the matrices ${\bar K}$, ${\bar L}$, ${\bar M}$ and ${\bar N}$
for the $SU(2)_J\times U(1)_Z\times U(1)_N$ projection are obtained by 
\begin{subequations}
\begin{eqnarray}
&&{\bar K}_{(k{\bar k}),\,(k^\prime \bar{k^\prime})}= 
{\bar L}_{(k{\bar k}),\,(k^\prime \bar{k^\prime})}^\ast 
\nonumber \\
&&\quad =\left(\begin{array}{cccc}
\bar{g}_k\xi_{kk^\prime}F_{kk^\prime}^\ast\bar{g}_{k^\prime} &
\bar{g}_k\xi_{kk^\prime}G_{k\bar{k^\prime}}^\ast\bar{g}_{\bar{k^\prime}} &
-\bar{g}_k\eta_{kk^\prime}G_{k\bar{k^\prime}}^\ast g_{k^\prime} &
\bar{g}_k\eta_{kk^\prime}F_{kk^\prime}^\ast g_{\bar{k^\prime}} \\ 
-\bar{g}_{\bar{k}}\xi_{kk^\prime}G_{k\bar{k^\prime}}\bar{g}_{k^\prime} &
\bar{g}_{\bar{k}}\xi_{kk^\prime}F_{kk^\prime}\bar{g}_{\bar{k^\prime}} &
-\bar{g}_{\bar{k}}\eta_{kk^\prime}F_{kk^\prime}g_{k^\prime} &
-\bar{g}_{\bar{k}}\eta_{kk^\prime}G_{k\bar{k^\prime}}g_{\bar{k^\prime}} \\
-g_k\eta_{kk^\prime}^\ast G_{k\bar{k^\prime}}\bar{g}_{k^\prime} &
g_k\eta_{kk^\prime}^\ast F_{kk^\prime}\bar{g}_{\bar{k^\prime}} &
g_k\xi_{kk^\prime}^\ast F_{kk^\prime}g_{k^\prime} &
g_k\xi_{kk^\prime}^\ast G_{k\bar{k^\prime}}g_{\bar{k^\prime}} \\
-g_{\bar{k}}\eta_{kk^\prime}^\ast F_{kk^\prime}^\ast\bar{g}_{k^\prime} &
-g_{\bar{k}}\eta_{kk^\prime}^\ast G_{k\bar{k^\prime}}^\ast
\bar{g}_{\bar{k^\prime}} &
-g_{\bar{k}}\xi_{kk^\prime}^\ast G_{k\bar{k^\prime}}^\ast g_{k^\prime} &
g_{\bar{k}}\xi_{kk^\prime}^\ast F_{kk^\prime}^\ast g_{\bar{k^\prime}}
\end{array}\right)  \nonumber \\
&&\qquad+~\delta_{kk^\prime}
\left(\begin{array}{cccc} g_k^2 & 0 & 0 & 0 \\ 
0 & g_{\bar{k}}^2 & 0 & 0 \\ 0 & 0 & \bar{g}_k^2 & 0 \\
0 & 0 & 0 & \bar{g}_{\bar{k}}^2 \end{array}\right),
\label{eq:eigseva} \\ 
&&{\bar M}_{(k{\bar k}),\,(k^\prime \bar{k^\prime})}= 
{\bar N}_{(k{\bar k}),\,(k^\prime \bar{k^\prime})}^\ast 
\nonumber \\
&&\quad =\left(\begin{array}{cccc}
-\bar{g}_k\eta_{kk^\prime}^\ast G_{k\bar{k^\prime}}\bar{g}_{k^\prime} &
\bar{g}_k\eta_{kk^\prime}^\ast F_{kk^\prime}\bar{g}_{\bar{k^\prime}} &
\bar{g}_k\xi_{kk^\prime}^\ast F_{kk^\prime}g_{k^\prime} &
\bar{g}_k\xi_{kk^\prime}^\ast G_{k\bar{k^\prime}}g_{\bar{k^\prime}} \\
-\bar{g}_{\bar{k}}\eta_{kk^\prime}^\ast F_{kk^\prime}^\ast\bar{g}_{k^\prime} &
-\bar{g}_{\bar{k}}\eta_{kk^\prime}^\ast G_{k\bar{k^\prime}}^\ast
\bar{g}_{\bar{k^\prime}} &
-\bar{g}_{\bar{k}}\xi_{kk^\prime}^\ast G_{k\bar{k^\prime}}^\ast g_{k^\prime} &
\bar{g}_{\bar{k}}\xi_{kk^\prime}^\ast F_{kk^\prime}^\ast
g_{\bar{k^\prime}} \\
g_k\xi_{kk^\prime}F_{kk^\prime}^\ast\bar{g}_{k^\prime} &
g_k\xi_{kk^\prime}G_{k\bar{k^\prime}}^\ast\bar{g}_{\bar{k^\prime}} &
-g_k\eta_{kk^\prime}G_{k\bar{k^\prime}}^\ast g_{k^\prime} &
g_k\eta_{kk^\prime}F_{kk^\prime}^\ast g_{\bar{k^\prime}} \\ 
-g_{\bar{k}}\xi_{kk^\prime}G_{k\bar{k^\prime}}\bar{g}_{k^\prime} &
g_{\bar{k}}\xi_{kk^\prime}F_{kk^\prime}\bar{g}_{\bar{k^\prime}} &
-g_{\bar{k}}\eta_{kk^\prime}F_{kk^\prime}g_{k^\prime} &
-g_{\bar{k}}\eta_{kk^\prime}G_{k\bar{k^\prime}}g_{\bar{k^\prime}}
\end{array}\right) \nonumber \\
&&\qquad -~\delta_{kk^\prime}
\left(\begin{array}{cccc} 0 & 0 & g_k\bar{g}_k & 0 \\
0 & 0 & 0 & g_{\bar k}\bar{g}_{\bar k} \\ 
\bar{g}_k g_k & 0 & 0 & 0 \\
0 & \bar{g}_{\bar k}g_{\bar k} & 0 & 0 \end{array}\right).
\label{eq:eigsevb}
\end{eqnarray}
\label{eq:eigsev}
\end{subequations}
For the case of the number projection only, putting $F=1$ and $G=0$ 
in the above expression and employing
$\xi_{kk}=\cos\varphi_\tau+i(u_k^2-v_k^2)\sin\varphi_\tau$,
$\eta_{kk}=-\eta_{kk}^\ast=2iu_kv_k\sin\varphi_\tau$,
we get 
\begin{subequations}
\begin{eqnarray}
{\bar K}_{(k{\bar k}),\,(k^\prime \bar{k^\prime})}= 
{\bar L}_{(k{\bar k}),\,(k^\prime \bar{k^\prime})}^\ast 
&=&\delta_{kk^\prime}\left(\begin{array}{cccc}
\bar{g}_k^2\xi_{kk}+g_k^2 & 0 & 0 & \bar{g}_k g_{\bar k}\eta_{kk} \\
0 & \bar{g}_{\bar k}^2\xi_{kk}+g_{\bar k}^2 & -\bar{g}_{\bar k}g_k\eta_{kk} &
 0 \\ 
0 & -g_k\bar{g}_{\bar k}\eta_{kk} & g_k^2\xi_{kk}^\ast+\bar{g}_k^2 & 0 \\ 
g_{\bar k}\bar{g}_k\eta_{kk} & 0 & 0 &
 g_{\bar k}^2\xi_{kk}^\ast+\bar{g}_{\bar k}^2 \end{array}\right),
\label{eq:eignina} \\ 
{\bar M}_{(k{\bar k}),\,(k^\prime \bar{k^\prime})}= 
{\bar N}_{(k{\bar k}),\,(k^\prime \bar{k^\prime})}^\ast 
&=&\delta_{kk^\prime}
\left(\begin{array}{cccc}
0 & -\bar{g}_k\bar{g}_{\bar k}\eta_{kk} & \bar{g}_k g_k(\xi_{kk}^\ast-1) &
 0 \\ 
\bar{g}_{\bar k}\bar{g}_k\eta_{kk} & 0 & 0 &
 \bar{g}_{\bar k}g_{\bar k}(\xi_{kk}^\ast-1)\\ 
g_k\bar{g}_k(\xi_{kk}-1) & 0 & 0 & g_k g_{\bar k}\eta_{kk} \\ 
0 & g_{\bar k}\bar{g}_{\bar k}(\xi_{kk}-1) & -g_{\bar k}g_k\eta_{kk} & 0
\end{array}\right). \nonumber\\
\label{eq:eigninb}
\end{eqnarray}
\label{eq:eignin}
\end{subequations}
Eq.~(\ref{eq:eignina}) yields
\begin{equation}
(\textrm{det}\bar{K})^{1/2}=\prod_{k>0}\,
\big[f_k(1-f_{\bar{k}})+(1-f_k)f_{\bar{k}}
+(1-f_k)(1-f_{\bar{k}})\xi_{kk}+f_k f_{\bar{k}}\xi_{kk}^\ast\big]\,.
\label{eq:eigninc}
\end{equation}
Eqs.~(\ref{eq:eignin},\ref{eq:eigninc}) give the formulae
equivalent to those in Ref.~\cite{EE93}.

\section{APPROXIMATION OF ENTROPY}
\label{sec:entropy}

Once the original Bogoliubov transformation (\ref{eq:one})
and the q.p. energy in Eq.~(\ref{eq:thr}) are determined,
the projection method discussed in Sec.~\ref{sec:proj} is
straightforwardly applicable.
This indicates the variation-before-projection (VBP) calculations.
However, there still remains a serious problem
in formulating the variation-after-projection (VAP)
calculations~\cite{RR94}.

Between the exact free energy $F^{\rm exact}$, 
which is generated by the exact Boltzmann-Gibbs operator 
$\exp(-{\hat H}/T)$, and 
an approximate one $F_\mathrm{P}$, which is generated 
by ${\hat P}\exp(-{\hat H}_0/T){\hat P}$ in our case, 
the Peierls inequality holds~\cite{TSM81}, 
\begin{equation}
F^{\rm exact}=-T\,{\rm ln}\big[{\rm Tr}
(e^{-{\hat H}/T}{\hat P})\big]
\leq F_\mathrm{P}=E_\mathrm{P}-TS_\mathrm{P}.
\label{eq:fotfiv}
\end{equation}
This inequality vindicates the variational calculations
minimizing $F_\mathrm{P}$.
The approximate energy $E_\mathrm{P}$ is expressed
in terms of the TFD vacuum expectation value by
\begin{equation}
E_\mathrm{P}={\rm Tr}({\hat w}_{\rm P}{\hat H})
=\frac{{\rm Tr}({\hat w}_0{\hat H}{\hat P})}
{{\rm Tr}({\hat w}_0{\hat P})}
=\frac{\langle 0_T|{\hat H}{\hat P}|0_T\rangle}
{\langle 0_T|{\hat P}| 0_T\rangle}. 
\label{eq:fotsix}
\end{equation}
The entropy is given by 
\begin{equation}
S_\mathrm{P}=-{\rm Tr}({\hat w}_{\rm P}{\rm ln}{\hat w}_{\rm P})
=-{\rm Tr}\Bigg[\frac{e^{-{\hat H}_0/T}{\hat P}}
{{\rm Tr}(e^{-{\hat H}_0/T}{\hat P})}
{\rm ln}\Bigg(\frac{e^{-{\hat H}_0/T}{\hat P}}
{{\rm Tr}(e^{-{\hat H}_0/T}{\hat P})}\Bigg)\Bigg], 
\label{eq:fotsev}
\end{equation}
where a relation ${\hat P}\ln({\hat P}{\hat A}{\hat P})
={\hat P}\ln({\hat A}{\hat P})$
for any operator ${\hat A}$ is applied.
However, no further reduction is allowed because 
$[{\hat P}, {\hat H}_0]\not=0$.
Since the projection operator remains in the logarithmic function,
it is prohibitively difficult to deal with
the entropy in Eq.~(\ref{eq:fotsev})
without further approximation.
 
In order to make the VAP calculations practical,
we here discuss an additional approximation on the entropy.
The VAP scheme is based on the fact that an approximate 
entropy does not exceed the exact one, 
or the Peierls inequality (\ref{eq:fotfiv}). 
Therefore, any approximation method 
adopted in the VAP scheme should not conflict 
with those inequalities.

We now introduce a projected space ${\cal P}$ 
and denote its complementary space by ${\cal Q}$.
The operator ${\hat P}e^{-{\hat H}_0/T}{\hat P}$
is hermitian and non-negative.
We here denote eigenstates of ${\hat P}e^{-{\hat H}_0/T}{\hat P}$
by $|K)$ and its eigenvalue by $\omega_K(\geq 0)$,
\begin{equation}
{\hat P}e^{-{\hat H}_0/T}{\hat P}\,|K)=|K)\,\omega_K\,.
\label{eq:nintwo}
\end{equation}
Any states in the $\mathcal{Q}$ are eigenstates of
${\hat P}e^{-{\hat H}_0/T}{\hat P}$ with the null eigenvalue.
The $\mathcal{P}$ space is also spanned
by eigenstates of ${\hat P}e^{-{\hat H}_0/T}{\hat P}$.
For $|K)\in\mathcal{P}$, $\hat{P}|K)=|K)$ leads to
\begin{equation}
\omega_K=(K|{\hat P}e^{-{\hat H}_0/T}{\hat P}|K)
=(K|e^{-{\hat H}_0/T}|K)\,.
\label{eq:ninsix}
\end{equation} 
Therefore, an essential part of the entropy is expressed as
\begin{equation}
{\rm Tr}\big[{\hat P}e^{-{\hat H}_0/T}{\hat P}
\ln({\hat P}e^{-{\hat H}_0/T}{\hat P})\big]
=\sum_{K\in{\cal P}}\omega_K\,\ln\omega_K 
=\sum_{K\in{\cal P}}(K|e^{-{\hat H}_0/T}|K)\,
\ln(K|e^{-{\hat H}_0/T}|K)\,.
\label{eq:ninsev}
\end{equation}
Observing that the functional form of the above 
expression is the convex function $F(x)=x\,\ln x$ with
$x=(K|e^{-{\hat H}_0/T}|K)$, we apply
the inequality (\ref{eq:btwo}) in Appendix B
to derive a useful inequality as follows,
\begin{eqnarray}
{\rm Tr}\big[{\hat P}e^{-{\hat H}_0/T}{\hat P}
\ln({\hat P}e^{-{\hat H}_0/T}{\hat P})\big]
&\leq& \sum_{K\in{\cal P}}(K|e^{-{\hat H}_0/T}
\ln e^{-{\hat H}_0/T}|K) \nonumber \\
&=&-\frac{1}{T} \sum_{K\in {\cal P}}
(K|e^{-{\hat H}_0/T}{\hat H}_0|K) \nonumber \\
&=&-\frac{1}{T} \sum_{K\in{\cal P}+{\cal Q}}(K|{\hat P}
e^{-{\hat H}_0/T}{\hat H}_0|K)
=-\frac{1}{T} \,{\rm Tr}(e^{-{\hat H}_0/T}{\hat H}_0{\hat P})\,.
\nonumber\\
\label{eq:nineig}
\end{eqnarray}
Thus, we obtain an inequality between two 
approximate entropies $S_\mathrm{P}$ and $S_\mathrm{P}^\prime$,
which is defined below,
\begin{eqnarray}
S_\mathrm{P}&=&-{\rm Tr}\Bigg[\frac{{\hat P}e^{-{\hat H}_0/T}
{\hat P}}{{\rm Tr}e^{-{\hat H}_0/T}{\hat P}}\,\ln
\Bigg(\frac{{\hat P}e^{-{\hat H}_0/T}{\hat P}}
{{\rm Tr}e^{-{\hat H}_0/T}{\hat P}}\Bigg)\Bigg]
\nonumber \\
&\geq& \frac{1}{T}\frac{{\rm Tr}(e^{-{\hat H}_0/T}
{\hat H}_0{\hat P})}
{{\rm Tr}(e^{-{\hat H}_0/T}{\hat P})}
+\ln{\rm Tr}(e^{-{\hat H}_0/T}{\hat P})
\equiv S_\mathrm{P}^\prime.
\label{eq:hundred}
\end{eqnarray}
The approximate free energy defined by $S_\mathrm{P}^\prime$
satisfies the Peierls inequality,
\begin{equation}
F_\mathrm{P}^\prime\equiv E_\mathrm{P}-TS_\mathrm{P}^\prime
\geq F_\mathrm{P}=E_\mathrm{P}-TS_\mathrm{P}
\geq F^\mathrm{exact}. 
\label{eq:hudone1}
\end{equation}

It is noted that this approximate entropy $S_\mathrm{P}^\prime$
can also be obtained by dealing with the logarithmic function
in Eq.~(\ref{eq:fotsev}) by $\hat{P}\ln(e^{-\hat{H}_0/T}\hat{P})\approx
-\hat{P}\hat{H}_0\hat{P}/T$,
as if ${\hat H}_0$ and ${\hat P}$ were commutable,
though such an ansatz is not adopted in the present argument.
An crucial point is that this approximation preserves
the Peierls inequality (\ref{eq:hudone1}), 
which gives a ground for the variational calculations.
Therefore, $F_\mathrm{P}^\prime$ will be available
for the VAP calculation,
since the smaller $F_\mathrm{P}^\prime$ gives
the better approximation to $F^\mathrm{exact}$.
The TFD expression of $F_\mathrm{P}^\prime$ is given by 
\begin{equation}
F_\mathrm{P}^\prime
=\frac{\langle 0_T|({\hat H}-{\hat H}_0){\hat P}|0_T\rangle}
{\langle 0_T|{\hat P}|0_T\rangle}
-T\,{\rm ln}\langle 0_T|{\hat P}|0_T\rangle
 +T\sum_{\mu({\rm all})}{\rm ln}(1-f_\mu).
\label{eq:hudone}
\end{equation}
In $F_\mathrm{P}^\prime$, all the quantities
can be calculated in the formalism presented in Sec.~\ref{sec:proj}.
For the $SU(2)_J\times U(1)_Z\times U(1)_N$ projection,
$\langle 0_T|{\hat P}|0_T\rangle$ and
$\langle 0_T|{\hat H}{\hat P}|0_T\rangle$
are given by Eq.~(\ref{eq:sevnin}) and Eq.~(\ref{eq:eigthr}),
respectively.
Moreover, $\langle 0_T|{\hat H}_0{\hat P}|0_T\rangle$
is immediately obtained from Eq.~(\ref{eq:ninone}).
Since these elements are expressed
as functions of the GBT coefficients
$U_{k\mu}$, $V_{k\mu}$, $U_{k\mu}^\ast$, $V_{k\mu}^\ast$
and $f_\mu=\sin^2\vartheta_\mu$ (\textit{i.e.} coefficients
of the TFD-extended GBT),
the variation of $F_\mathrm{P}^\prime$ with respect to these variables
produces a closed set of independent VAP equations. 

\section{SUMMARY and DISCUSSIONS}
\label{sec:summary}

Although the mean-field theories play an important role
in nuclear physics at finite temperature as well as at zero temperature,
the mean-field solutions often violate conservation laws.
In order to restore desired quantum numbers
for thermally equilibrated many-body states of 
an isolated finite system like a nucleus, 
we have applied the thermo field dynamics (TFD) 
to construct a transparent and practical formalism 
of the quantum-number-projected statistics. In the TFD, 
an ensemble average of an observable is expressed 
by a TFD vacuum expectation value,
whereas the single-particle operator space 
is doubled by introducing tilded operators. 
Making use of this method, 
we have derived formulae for the projected statistics
at finite temperature. 
As an example, we have shown explicit representations
for the projection of angular momentum
as well as that of particle numbers.
A significant advantage of this formalism is 
to keep a complete parallel with the zero-temperature case,
so that the projection method and the 
corresponding computer code, which have been demonstrated 
to be successful, could be directly translated to the 
projection at finite temperature, apart from the variation. 

Our formalism presented in Sec.~\ref{sec:proj} is, in principle, 
applicable to both the variation-before-projection (VBP)
and the variation-after-projection (VAP) schemes.
In the mean-field calculations without the projections,
sharp phase transitions appear;
for instance, there could be a discontinuity
in heat capacity at the critical temperature.
This is not realistic for finite systems,
in which quantum fluctuations of the fields often wash out
the distinct signature of transitions.
While such unrealistic signature remains
by the VBP calculations,
the VAP scheme is expected to smooth it out,
including a significant portion of the fluctuations.
However, the VAP calculations minimizing the projected free energy
are impractical without further approximation,
since the entropy does not take a simple form
because of the non-commutability of ${\hat P}$ with ${\hat H}_0$.
Thus, in Sec.~\ref{sec:entropy}, 
we have discussed an additional approximation for the entropy,
which is expected to resolve this problem.
It should be emphasized that the Peierls inequality,
which is an important requisite justifying the variation,
is proven to be satisfied in this approximation.

We have adopted the HFB or constrained HFB (CHFB)
self-consistent solution at finite temperature
for the basic quasiparticle picture,
and have developed the TFD formalism on it.
The CHFB approximation is the most effective mean-field theory
in describing nuclear structure at low temperature or along the yrast.
It is commented here that,
whereas the constraints might not look important
when the projection is implemented, it can still be useful.
In the VBP calculation, it could be essentially important
to obtain a mean-field solution well connected
to the true many-body solution.
Also in the VAP calculations, the CHFB solution
may be used to obtain good initial configurations.
Furthermore, if there remain some quantum numbers unprojected,
${\hat C}_j$ $(j=1, 2, \cdots, N_c)$,
they could be handled in the CHFB scheme
in the projected statistics for other quantum numbers,
by replacing the Hamiltonian ${\hat H}$ by the auxiliary one 
${\hat H}^\prime={\hat H}-\sum_{j=1}^{N_c}\lambda_j{\hat C}_j$ in the 
free energy in Eqs.~(\ref{eq:fotfiv},\ref{eq:fotsix},\ref{eq:hudone}). 

The present formalism will be useful
in investigating effects of quantum fluctuations 
connected to the conservation laws on thermal properties of nuclei.
A topical example is the pairing correlations at finite temperature.
Although the superfluid-to-normal phase transition
has been predicted to occur at relatively low temperature
($0.5\lesssim T_c\lesssim 1\,\textrm{MeV}$)~\cite{TS80,Goo86,ERM00}, 
it is not yet well established in nuclei.
Based on a precise measurement of nuclear level densities,
the superfluid-to-normal transition in nuclei
has been discussed recently~\cite{Sch01}.
While there is no sharp discontinuity as in infinite systems,
the $S$-shape behavior in the graph
of heat capacity \textit{vs.} temperature, $C(T)$,
has been argued to be a signature of the transition~\cite{Sch01,LA01}.
For proper understanding of the phase transition
and its relation to the $S$-shape,
it is important to investigate effect of the quantum 
fluctuations in connection with the conservation laws, 
although the other quantum fluctuations cannot be neglected 
in quantitative description~\cite{RCR98}, particularly around 
the critical temperature. This situation holds also 
for the deformed-to-spherical shape phase transition.
 
Several theoretical frameworks have been invented
to take into account the quantum fluctuations 
not restricted to those connected to the conservation laws; 
for instance, the static-path approximation
(SPA)~\cite{LA84,AZ84,AB88} 
and the shell-model Monte Carlo (SMMC) approach~\cite{LJKO93}. 
Whereas we have mainly focused our discussion on the projections 
in the mean-field approximations, 
it will be straightforward to incorporate the projections 
into the above extensive methods, 
in which the wave functions are represented
by a superposition of those in the mean-field theories.
In practice, the SPA calculations with the particle-number
and angular momentum projections
have been carried out~\cite{RR94,RAR93},
based on the conventional thermal formalism.
With its simplicity the present TFD formulation of projections
will be useful also in this course.
The projections combined with the SMMC have been restricted
to relatively simple cases~\cite{NA97,LJKO93},
in which $[\hat{P},\hat{H}_0]=0$ holds.
The present formulation may help applying
more general projections to the SMMC implementation;
\textit{e.g.} the angular momentum projection,
and the particle-number projection in the pairing decomposition.

\begin{acknowledgments}

One of the authors (H. N.) acknowledge the financial support of 
the Grant-in-Aid for Scientific Research (B), No.~15340070, 
by the Ministry of Education, Culture, Sports, Science and Technology, 
Japan. 
\end{acknowledgments}

\appendix

\section{Quasiparticle representation of Hamiltonian} 

We here present the quasiparticle representation of the Hamiltonian
$\hat{H}^\prime$ given in Eq.~(\ref{eq:fiv}).
The Hamiltonian is assumed to consist of up to the two-body interactions
as in Eq.~(\ref{eq:nin}).
Applying the Bogoliubov transformation in Eq.~(\ref{eq:one}),
we rewrite $\hat{H}^\prime$ in terms of quasiparticle operators,
\begin{subequations}
\begin{eqnarray}
U_0&=&\langle\mbox{vac.}|{\hat H}^\prime|\mbox{vac.}\rangle
-\lambda_{\rm p}Z-\lambda_{\rm n}N-\omega_{\rm rot}\sqrt{J(J+1)}
\nonumber \\
&=&{\rm Tr}_\mathrm{s.p.}\bigg(\xi\rho^{(0)}
+\frac{1}{2}\Gamma^{(0)}\rho^{(0)}
+\frac{1}{2}\Delta^{(0)}\kappa^{(0)\dagger}\bigg), \label{eq:aonea} \\
{\hat H}_{11}&=&\sum_{\mu\nu}(H_{11})_{\mu\nu}
\alpha_\mu^\dagger\alpha_\nu, \label{eq:aoneb} \\
{\hat H}_{20}&=&\frac{1}{2}\sum_{\mu\nu}\big[(H_{20})_{\mu\nu}
\alpha_\mu^\dagger\alpha_\nu^\dagger+(H_{20})_{\mu\nu}^\ast
\alpha_\nu\alpha_\mu\big], \label{eq:aonec} \\
{\hat H}_{22}&=&\sum_{\mu\nu\rho\sigma}
(H_{22})_{\mu\nu\rho\sigma}\alpha_\mu^\dagger
\alpha_\nu^\dagger\alpha_\sigma\alpha_\rho, \label{eq:aoned} \\
{\hat H}_{31}&=&\sum_{\mu\nu\rho\sigma}\big[
(H_{31})_{\mu\nu\rho\sigma}\alpha_\mu^\dagger\alpha_\nu^\dagger
\alpha_\rho^\dagger\alpha_\sigma
+(H_{31})_{\mu\nu\rho\sigma}^\ast\alpha_\sigma^\dagger
\alpha_\rho\alpha_\nu\alpha_\mu\big], \label{eq:aonee} \\
{\hat H}_{40}&=&\sum_{\mu\nu\rho\sigma}\big[
(H_{40})_{\mu\nu\rho\sigma}\alpha_\mu^\dagger\alpha_\nu^\dagger
\alpha_\rho^\dagger\alpha_\sigma^\dagger
+(H_{40})_{\mu\nu\rho\sigma}^\ast \alpha_\sigma\alpha_\rho
\alpha_\nu\alpha_\mu\big], \label{eq:aonef} 
\end{eqnarray}
\end{subequations}
where 
\begin{subequations}
\begin{eqnarray}
(H_{11})_{\mu\nu}&=&\big[U^\dagger(\xi+\Gamma^{(0)})U
-V^\dagger(\xi+\Gamma^{(0)})^\ast V
 +U^\dagger\Delta^{(0)} V-V^\dagger\Delta^{(0)\ast} U\big]_{\mu\nu}, 
\label{eq:atwoa} \\
(H_{20})_{\mu\nu}&=&\big[U^\dagger(\xi+\Gamma^{(0)})V^\ast
-V^\dagger(\xi+\Gamma^{(0)})^\ast U^\ast
+U^\dagger\Delta^{(0)} U^\ast
-V^\dagger\Delta^{(0)\ast} V^\ast\big]_{\mu\nu}, \label{eq:atwob} \\
(H_{22})_{\mu\nu\rho\sigma}&=&\frac{1}{4}\sum_{ijkl}v_{ijkl}
\big[U_{i\mu}^\ast U_{j\nu}^\ast U_{k\rho}U_{l\sigma}
+V_{k\mu}^\ast V_{l\nu}^\ast V_{i\rho}V_{j\sigma}
-U_{i\mu}^\ast V_{l\nu}^\ast U_{k\rho}V_{j\sigma} 
+U_{i\nu}^\ast V_{l\mu}^\ast U_{k\rho}V_{j\sigma} \nonumber\\
&&\hspace*{2cm}+U_{i\mu}^\ast V_{l\nu}^\ast U_{k\sigma}V_{j\rho} 
-U_{i\nu}^\ast V_{l\mu}^\ast U_{k\sigma}V_{j\rho}\big], 
\label{eq:atwoc} \\ 
(H_{31})_{\mu\nu\rho\sigma}&=&\frac{1}{6}\sum_{ijkl}v_{ijkl}
\big[U_{i\mu}^\ast U_{j\nu}^\ast U_{k\sigma}V_{l\rho}^\ast
+U_{i\mu}^\ast V_{j\sigma}V_{k\rho}^\ast V_{l\nu}^\ast
+U_{i\nu}^\ast U_{j\rho}^\ast U_{k\sigma}V_{l\mu}^\ast
+U_{i\nu}^\ast V_{j\sigma}V_{k\mu}^\ast V_{l\rho}^\ast \nonumber\\
&&\hspace*{2cm}+U_{i\rho}^\ast U_{j\mu}^\ast U_{k\sigma}V_{l\nu}^\ast
+U_{i\rho}^\ast V_{j\sigma}V_{k\nu}^\ast V_{l\mu}^\ast\big],
\label{eq:atwod} \\
(H_{40})_{\mu\nu\rho\sigma}&=&\frac{1}{24}\sum_{ijkl}v_{ijkl}
\big[U_{i\mu}^\ast U_{j\nu}^\ast V_{k\rho}^\ast V_{l\sigma}^\ast
+U_{i\rho}^\ast U_{j\sigma}^\ast V_{k\mu}^\ast V_{l\nu}^\ast
-U_{i\rho}^\ast U_{j\nu}^\ast V_{k\mu}^\ast V_{l\sigma}^\ast
-U_{i\mu}^\ast U_{j\sigma}^\ast V_{k\rho}^\ast V_{l\nu}^\ast \nonumber\\
&&\hspace*{2cm}-U_{i\sigma}^\ast U_{j\nu}^\ast V_{k\rho}^\ast V_{l\mu}^\ast
-U_{i\mu}^\ast U_{j\rho}^\ast V_{k\sigma}^\ast V_{l\sigma}^\ast\big]
\,. \label{eq:atwoe} 
\end{eqnarray}
\label{eq:atwo}
\end{subequations}
Definition of $\xi$ is the same as given in Eq.~(\ref{eq:thirt}).
$\Gamma^{(0)}$ and  $\Delta^{(0)}$ are defined
in an analogous manner to Eq.~(\ref{eq:eight}),
with the single-particle density $\rho^{(0)}$ and 
the pair tensor $\kappa^{(0)}$ defined by 
\begin{subequations}
\begin{eqnarray}
\rho^{(0)}_{ij}&=&(\rho^{(0)\dagger})_{ij}
=\langle\mbox{vac.}|c_j^\dagger c_i|\mbox{vac.}\rangle
=(V^\ast V^{\rm tr})_{ij}\,, \label{eq:athra} \\
\kappa^{(0)}_{ij}&=&-\kappa^{(0)}_{ji}
=\langle\mbox{vac.}|c_j c_i|\mbox{vac.}\rangle
=(V^\ast U^{\rm tr})_{ij}\,. \label{eq:athrb}
\end{eqnarray}
\label{eq:athr}
\end{subequations}

\section{PROOF OF INEQUALITY used in section~\ref{sec:entropy}} 

\par
Let $F(x)$ be a real function, 
which is convex downwards (or 
concave upwards) in a simply connected domain of 
the variable $x$ $(>0)$, and $x_0$ belongs to the same domain.
This leads to the inequality 
\begin{equation}
F(x)\geq F(x_0)+F^\prime (x_0)(x-x_0)\,.
\label{eq:bone}
\end{equation}
As far as $F''(x)>0$, the the left-hand side (lhs)
and the right-hand side (rhs) of Eq.~(\ref{eq:bone})
become equal only at $x=x_0$.
For a hermitian and non-negative operator ${\hat O}$
and a given physical state $|i\rangle$,
the following relation turns out,
\begin{equation}
\langle i|F({\hat O})|i\rangle\geq 
F(\langle i|{\hat O}|i\rangle)\,.
\label{eq:btwo}
\end{equation}
This inequality is proven as follows.

First, we introduce a complete set of the eigenstates of 
the operator ${\hat O}$, denoted by $\lbrace|k\rangle\rbrace$.
Each eigenstate $|k\rangle$ satisfies
\begin{equation}
{\hat O}|k\rangle=|k\rangle\,\omega_k. 
\label{eq:bthr}
\end{equation}
Then, the state $|i\rangle$ is expanded by $\lbrace|k\rangle\rbrace$,
\begin{equation}
|i\rangle=\sum_k|k\rangle\, u_{ki},
\label{eq:bfou}
\end{equation}
which is a unitary transformation.
Using Eqs.~(\ref{eq:bone},\ref{eq:bfou}),
we obtain
\begin{eqnarray}
\langle i|F({\hat O})|i\rangle&=&
\sum_k|u_{ki}|^2\,F(\omega_k) \nonumber \\
&\geq& \sum_k|u_{ki}|^2\,\big[F(\langle i|{\hat O}|i\rangle) 
+F^\prime(\langle i|{\hat O}|i\rangle)(\omega_k
-\langle i|{\hat O}|i\rangle)\big]
=F(\langle i|{\hat O}|i\rangle), 
\label{eq:bfiv}
\end{eqnarray}
by applying Eq.~(\ref{eq:bone}) with setting $x=\omega_k$
and $x_0=\langle i|{\hat O}|i\rangle$, and using
$\sum_k|u_{ki}|^2=1$ and 
$\sum_k|u_{ki}|^2\,\omega_k=\langle i|{\hat O}|i\rangle$.
It is now obvious that the lhs and the rhs of Eq.~(\ref{eq:btwo})
are equal only if $|i\rangle$ is an eigenstate of $\hat{O}$.

It has been verified~\cite{Pei38} that,
if $F'(x)<0$ and $F''(x)>0$, we have
\begin{equation}
\sum_i \langle i|F({\hat O})|i\rangle\geq 
\sum_i F(\langle i|{\hat O}|i\rangle)\,.
\label{eq:bsix}
\end{equation}
We have here derived the inequality (\ref{eq:btwo})
in more general manner,
by lifting the sum over $i$ and the condition $F'(x)<0$,
besides that truncation scheme is also discussed in Ref.~\cite{Pei38}.

 
\end{document}